# Dust aerosol, clouds, and the atmospheric optical depth record over 5 Mars years of the Mars Exploration Rover mission

Mark T. Lemmon[a], Michael J. Wolff[b], James F. Bell III[c], Michael D. Smith[d], Bruce A. Cantor[e], and Peter H. Smith[f]

[a] Department of Atmospheric Sciences, Texas A&M University, College Station, Texas 77843-3150, USA; lemmon@tamu.edu

[b] Space Science Institute, Boulder, Colorado, USA

[c] School of Earth and Space Exploration, Arizona State University, Tempe, Arizona, USA

[d] NASA Goddard Space Flight Center, Greenbelt, MD, USA

[e] Malin Space Science Systems, San Diego, California, USA

[f] Department of Planetary Sciences, University of Arizona, Tucson, Arizona, USA

Corresponding author: Mark Lemmon, Dept. of Atmospheric Sciences, Texas A&M University, TAMU 3150, College Station, TX 77843-3150. E-mail: lemmon@tamu.edu. Phone: 1-979-862-8098, Cell: 1-979-777-2831, Fax: 1-979-862-4466.

## Highlights

- The derivation of the Mars Exploration Rover opacity record is described
- Dust aerosol size variations are characterized across seasons and storm events
- Clouds contribute to northern summer optical depth at the Opportunity site
- The dust significantly affects the energy balance and frequency of dust devils

# Abstract

Dust aerosol plays a fundamental role in the behavior and evolution of the Martian atmosphere. The first five Mars years of Mars Exploration Rover data provide an unprecedented record of the dust load at two sites. This record is useful for characterization of the atmosphere at the sites and as ground truth for orbital observations. Atmospheric extinction optical depths have been derived from solar images after calibration and correction for time-varying dust that has accumulated on the camera windows. The record includes local, regional, and globally extensive dust storms. Comparison with contemporaneous thermal infrared data suggests significant variation in the size of the dust aerosols, with a 1 μm effective radius during northern summer and a 2 μm effective radius at the onset of a dust lifting event. The solar longitude ($L_S$) 20-136° period is also characterized by the presence of cirriform clouds at the Opportunity site, especially near $L_S$=50 and 115°. In addition to water ice clouds, a water ice haze may also be present, and carbon dioxide clouds may be present early in the season. Variations in dust opacity are important to the energy balance of each site, and work with seasonal variations in insolation to control dust devil frequency at the Spirit site.

# 1. Introduction

Dust aerosol plays a fundamental role in the behavior and evolution of the Martian atmosphere. Atmospheric dust and ice clouds have a direct effect on both surface and atmospheric heating rates, which are basic drivers of atmospheric dynamics. In turn, the resulting atmospheric motions influence the distribution of the dust itself, thus setting up a complex feedback [*e.g.,* Newman *et al.,* 2002]. For example, the lifting of dust from the surface of Mars is a function of winds and atmospheric instability (*i.e.,* turbulence) at and near the surface, both of which are influenced in turn by heating rates dependent of the amount of atmospheric dust loading. This type of non-linear feedback mechanism is suspected of determining whether in a given year Mars will see a global dust event or only an assortment of local to regional storms [*e.g.,* Rafkin, 2009]. In addition, the changes in the dust column can directly influence the vertical scale of the atmosphere, and sometimes produce large variations in density (observed in the form of aerodynamical drag) at orbital and aerobraking altitudes [Bell *et al.,* 2007].

The importance of dust was recognized early in the exploration of Mars [Gierasch and Goody, 1972] and consequently dust became a target of atmospheric studies [see review by Smith, 2008]. While earth-based and orbital observations comprise a large part of the historical record, landed observations of optical depth represent a critically important component because of the higher accuracy and precision generally available from such measurements. That is to say, although orbital observations allow for a more complete characterization of dust loading over spatial and temporal scales, they generally embody more modeling or retrieval assumptions than one might find with surface observations [*e.g.,* Smith, 2004; Smith, 2009; Wolff *et al.* 2009]. In essence, they cannot provide generally the self-consistent "ground-truth", the proverbial "grail" of remote sensing algorithm validation. Prior to the Mars Exploration Rover (MER) mission, the record of surface observations comprised about 1.3 Martian years of data at each of the two Viking Lander sites [Colburn *et al.,* 1989] and about 0.1 Martian year of data at the Pathfinder site [Smith and Lemmon, 1999]. However, despite this temporal coverage, there were no simultaneous orbiter/ground station observations. In fact, even the application of these data to close comparisons has been fairly limited [Hunt, 1979; Martin, 1986; Clancy and Lee, 1991].

In January 2004, a new era of surface-based atmospheric observations began with the arrival of the MER missions. The two rovers, MER-A (Spirit) and MER-B (Opportunity) landed shortly before the Southern autumn equinox on opposite sides of the planet in Gusev crater (14.57°S, 175.48°E, -1.9 km) and in Meridiani Planum (1.95°S, 354.47°E, -1.8 km), respectively [Arvidson *et al.,* 2006; Squyres *et al.,* 2006]. Although "mission success" was considered to be 90 sols (a sol is a Martian solar day, about 1.0275 Earth solar days) of operation for each rover, Spirit functioned through sol 2209 and Opportunity has accumulated more than 3500 sols as of this writing. Throughout their mission, the rovers have provided atmospheric optical depth measurements though direct solar imaging [Lemmon *et al.*, 2004]. The evolving optical depth data set has been used in coordinated ("ground truth") and contextual observations of the atmosphere and surface [*e.g.,* Wolff *et al.,* 2006, 2009, 2010; Mateshvili *et al.,* 2007; Clancy *et al*., 2007; Vincendon *et al.,* 2007, 2009; Heavens *et al.,* 2011; Hoekzema *et al*., 2011; Petrova *et al*., 2012; Montabone *et al*., this issue]. In addition, this dataset has been applied to investigations of surface activity [*e.g*., Arvidson *et al*., 2006; Greeley *et al*., 2006a, 2006b; Rice *et al*., 2011], photometric models investigating surface or atmospheric properties [*e.g*., Johnson *et al.,* 2006a, 2006b; Bell *et al*., 2006b; Kinch *et al*., 2007; Shaw *et al*., 2013], and as inputs to discussions of chemical properties or biopotential [*e.g*., Schuerger *et al.,* 2006, 2008; Vincendon *et al*., 2010]. Although the MER solar observations were briefly described by Lemmon *et al*. [2004], the scope and the complexity of the current dataset have greatly exceeded that presentation. In this paper, we describe in detail the generation of MER optical depth record and the associated uncertainties, as necessary to understand the archival data. Section 2 describes the observations, their operational context, and the on-board processing done by the rovers. Section 3 describes the processing of the downlinked data into an optical depth archive. Section 4 presents a discussion of implications of the data set, and section 5 presents the conclusions.

## 2. Observations

### 2.1. Overview and rover operations context

The MER atmospheric optical depth archive spans more than 2200 sols for Spirit and more than 3500 sols for Opportunity. The mission began near the end of Mars year (MY) 26, using the Clancy *et al*. [2000] convention for numbering. In this work, we consider the entire Spirit data set and the Opportunity data set through the northern spring equinox at the end of MY 31 (31 July 2013). Pancam solar images using two neutral-density (ND) coated filters form the fundamental measurements. We first describe the imaging and raw Experiment Data Records (EDRs) in this section; details of the optical depth retrieval are provided in the next section.

For context, the overall pattern of optical depth is shown in Fig. 1, and is not strongly sensitive to the calibration details described below for sols 1-1200. Seasonal context is discussed in section 4, but generally low and stable opacities signify southern winter (and aphelion), while high and variable opacities correlate with southern summer (and perihelion).

Pancam has two cameras (eyes) per rover, each with a CCD detector, a filter wheel, and optics that map a 16°x16° field of view onto 1024x1024 pixels [Bell *et al.*, 2003]. The solar filters are filter number eight in each of the right and left Pancams. The nominal wavelengths are 440 nm (L8) and 880 nm (R8); comparable Viking observations were done at 670 nm [Colburn *et al.,* 1989], and Pathfinder observations were made at 440, 670, 880, and 990 nm [Smith and Lemmon, 1999]. Late in pre-launch calibration, a red-leak—excess long wavelength sensitivity—was discovered in the L8 filters. This defect was not significant prior to application of the ND coating, and was imprecisely measured after the coating due to low signal. The leak affects a broad wavelength range, roughly centered at 750 nm. The red leak is somewhat more significant in the pre-flight calibration, using a cooler lamp, than in flight, imaging a hot Sun. Nonetheless, in general we refer to the optical depths by filter name, rather than wavelength.

Most solar images were taken as part of a dedicated campaign to create a detailed optical depth record. Additional images were also obtained in order to characterize the optical depth as context for another primary observation, such as sky or surface photometry. In addition, the Sun's apparent motion over time was used by software aboard the rovers to determine spacecraft attitude after some drives. A subset of the latter ("engineering") images is included in this analysis. A sequence name of P26xx (*e.g.,* P2601) indicates an atmospheric-science sequence; sequences named P27xx was used for some lunar transit imaging; and F0006 designates an "engineering" sequence. An example product identifier is 2P127616614ESF0309P2600R8M1, where '2P' indicates Spirit's Pancam, there is a 9-digit time stamp, a product type (ESF = subframe, ETH = thumbnail image), a 4-character location identifier, sequence (P2600), filter (R8), a producer code and version number. The product identifier is included in the archival record described in the next section.

One of the standard monitoring sequences was routinely done shortly before noon (specifically immediately before or after the transition from one sol's master sequence to the next) to maximize completeness of a record for monitoring performance of the solar arrays. All solar images were exposed for a commanded duration rather than using auto-exposure; imaging when the Sun was low or the optical depth was high required longer exposures that are typical of some (e.g., P2633) sequences. Observations were commanded generally as paired L8-R8 exposures. Exceptions include some Phobos and Deimos transit observations (R8 only) and F0006 sequences (L8 only). A somewhat larger subset of the observations in the record are not paired. Reasons include: differences in exposure times resulting in one image saturating or under-exposing when conditions rapidly change; failure in the on board Sun detection, sometimes due to a faint Sun and a bright cosmic ray; and unrecovered loss of data during transmission.

Figure 2 shows the time-of-sol sampling for each rover. The nominal operational day for the rover was initially about 1100 to 1700 HLST (hybrid local solar time—this is the local mean solar time, offset to make it close to local true solar time during the nominal mission). On certain sols, imaging earlier and later was possible, particularly when solar power levels were high (*e.g.*, Southern summer and/or clean solar arrays). Early-morning imaging was especially rare due to the low temperatures and the need to heat actuators and electronics prior to use. In addition, frequent optical depth observations were made early in the mission; as mission longevity raised concerns about actuator lifetime, the number of observations per sol was scaled back. Patterns seen in Fig. 2 reflect the operational part of the sol and the tendency to fix opacity measurements to communications windows: direct to/from Earth communications near 11 HLST, and Mars Odyssey overflights near 17 LTST (local true solar time, defined by local true midnight, and varying in sol duration due to Mars' eccentric orbit). The figure also illustrates the changing relation between HLST and LTST (the equation of time) in the pattern oscillating around 12 LTST.

The procedures and data products described in this paper were developed in conjunction with evolving operations concepts. The amount of data collected and the specifics of the implementation varied with lessons learned and in response to operational pressures. A key focus of the process was to keep a quick-look optical depth available on a rapid timescale, while maintaining consistency with the final product.

### 2.2. Rover on-board processing and downlink

A typical solar-image command used the rover's on board ephemeris to aim the camera at the Sun. This approach allowed for flexibility with respect to the time a sequence is run (*e.g.*, P2600 was run at many times of sol, independently of rover attitude or the Sun's position), but was susceptible to drift in the Sun's position in the image frame when a rover's attitude self-knowledge is degraded. Attitude knowledge was typically good to within 3°, and was restored with dedicated solar imaging (F0006 images) and Mini Inertial Measurement Unit (MIMU) use based on operational needs. The typical solar-imaging command also resulted in 4 EDRs being returned to Earth, a subframed image and a thumbnail image in each of the two filters. "Sunfind", or auto-subframed, images, are 63x63 pixels from the original image, centered on the ~20-pixel diameter Sun; manual subframes were required to be larger parts of the original image, to account for pointing and timing uncertainties. Thumbnail images were automatically created for all MER images and "downlinked" with high priority for quick operational assessment of a sequence's results. These thumbnail images were available before the subframes (sometimes by weeks), and in some cases remain the only products available (*e.g.*, due to failure of the automatic subframing or loss of data).

Exposure durations were typically 3-5 s for L8 and 0.3-0.4 s for R8, but exposures of up to 5 minutes were possible. However, exposure times of over 1 minute "trail" the Sun by at least its own diameter and were avoided. Pancam's image acquisition relies on an electronic, rather than physical, shutter. That is to say, image readout is continuous until the start of an acquisition, when readout pauses for the exposure duration and then a readout is recorded. A zero-second ("shutter") exposure would see only a shutter smear caused by the 5 ms readout while the detector is exposed to light. Longer exposures would superimpose the image on that smear. The smear could be corrected out at image acquisition by subtracting an actual zero-second image or after downlink by analytically synthesizing a shutter image. Due to the Sun's apparent motion between actual exposures, the synthetic shutter method was typically relied on; operational issues prevented that mode from being used exclusively.

To minimize data volume, solar images were subframed to be as small as practical. The automated subframing algorithm relied upon a digital filter applied to the stored image, looking for the maximum brightness of a central box slightly larger than the Sun after subtracting the average of the pixels adjacent to the box. If the summed data numbers (DN) exceeded some threshold, a 63x63 subframe was extracted and stored for downlink. In order to image the Sun when the predicted brightness might fail to meet the threshold, manual subframes were used. Such cases included low Sun and high optical depth situations when even long exposures would be faint, or when the expected variability was sufficiently high that the range of possible brightness was too large for the range between the threshold and detector saturation. Images were sometimes companded from 12- to 8-bits per pixel via on board look-up table and were generally losslessly compressed.

Thumbnail images, 64x64 images that have been downsampled from 1024x1024 full frame images using pixel averaging, then compressed to 4 kbit each, were created for all MER images and downlinked at higher priority than any regular images. As such, the solar-image thumbnails formed the basis for a quick-look optical depth, being processed to optical depth in the same manner as the full-resolution images. In the cases where they are the only record, the thumbnail data are merged into the archive.

A lesson for operation of future missions is that the metric for total DN within a local-background-subtracted sun-sized box is a meaningful number. This single number is roughly proportional to flux, subject to a few perturbations described below. The on-board processing was one step away from downlinking an operationally useful quick-look optical depth as a single number attached to an image label, with no actual image needed. The initial build of flight software returned this flux-like number for even failed sunfind attempts (*e.g.,* when the Sun was detectable but failed to meet the threshold), thus allowing optical depth retrieval from the image header even when the conservatively set threshold caused rejection of an otherwise valid image. The intensity gap in which images could be processed to provide a useful optical depth measurement yet fail to meet the threshold is due to an operationally necessary compromise. Science use of the data would motivate a low threshold in order to gather the most data; however, engineering use required a high threshold in order to avoid false positives influencing attitude derivation, and only a single threshold was available.

## 3. Optical depth calibration

### 3.1. Overview

Conversion of solar images into an optical depth measurement is fundamentally simple: following the Beer-Lambert law, $F = F_{TOA} \cdot e^{-\tau \cdot \eta}$, where $F$ is the measured flux, $F_{TOA}$ is the top-of-atmosphere flux, $\tau$ is the desired normal-incidence extinction optical depth, and $\eta$ is the airmass, or ratio of slant-path optical depth to normal optical depth. Significant variations in the true $F_{TOA}$ varies over time inversely to the square of the known Mars-Sun distance. Figure 3 shows example solutions using subsets of the data. The trick is in the accurate determination of $F$ and $F_{TOA}$ in self-consistent units to minimize systematic errors, and to a lesser extent in the determination of $\eta$. The methods employed are described in this section.

Derivation of atmospheric transmittance, $F/F_{TOA}$, was done via a relative calibration approach similar to that in the Mars Pathfinder analysis of Smith and Lemmon [1999]. Although the absolute calibration of the L8 and R8 filter response was sufficient to determine exposure times, comparison of an instrumentally calibrated flux with a "known" solar flux would result in errors that were large compared to the amplitude of interesting atmospheric variability, and specifically would introduce spurious diurnal variability controlled by airmass. Ideally, $F_{TOA}$ would be directly observed, but this was impossible. Instead, one may take advantage of the linearity of $\ln(F) = \ln(F_{TOA}) - \tau \cdot \eta$. By asserting that $\tau$ is unchanging in a period containing several measurements at different η, the slope ($\tau$) and intercept ($\ln((F_{TOA}))$) can be retrieved. Complications associated with this methodology arise from systematic and random natural variability, and from the fact that (by the end of the observational record) the cameras had spent 9 years in a dusty environment. We discuss specific solutions below.

Image calibration for optical depth retrieval was done via a dedicated pipeline. Image calibration was done outside of the standard calibration pipeline [Bell *et al*. 2006a] for the following reasons. (1) There was a mix of images with and without onboard shutter subtraction, requiring those without to get synthetic shutter correction for consistency. In general, subframed images cannot and did not have a synthetic shutter generated. In the special case when the subframe contains the brightest object around by orders of magnitude, the generality does not hold. (2) Calibration must subtract a dark current and bias value from the images. The standard pipeline did not report negative intensities, effectively adding flux to the background when the subtraction was overdone. Accurate background correction is important, as described below. (3) It quickly became apparent during operations that control over the temperature and flat field corrections would help to produce more accurate results.

### 3.2. Image processing

A raw image experiment data record for a subframe was expanded from 8 to 12 bits via inverse lookup table when needed. The image was then calibrated using elements of the Pancam calibration pipeline [Bell et al., 2006a], following the Bell et al. [2006a] flow from image input to flat-field removal. The procedure involved subtracting out bias and dark current and the temperatures and exposure time reported in the image header; subtracting a model shutter image if needed; then dividing out the exposure time. Thus, $F_{i,j} = (S(i,j) - B(T) - D(i,j,T) \cdot t_{exp} - Sh(i,j)) / t_{exp}$, where $F_{i,j}$ is the derived flux for pixel $i,j$ in DN/s; $S$ is the observed signal in DN, $B$ is the modeled bias at temperature $T$, $D$ is the modeled dark current rate, and exposure duration $t_{exp}$, and $Sh$ is the modeled 0-sec shutter or readout image.

Flux extraction required finding the Sun as the brightest pixel in a boxcar smoothed version of the image (a small number of false positives were manually screened). The Sun's position was then determined more precisely via the centroid in the original image—note that this position was almost always the image center for auto-subframed images. The background flux was determined in an annulus ~20-30 pixels (2-3 solar radii) from the center. The residual background came from a combination of error in bias and dark subtraction and the sometimes-detectable, forward-scattered diffuse light. The influence of scattered light increases with airmass, so the last step was needed to avoid systematic errors. After background subtraction, pixels were summed to retrieve a flux in DN/s and normalized to a constant Sun-Mars distance of 1.5 AU. Images with saturated pixels were flagged with a negative flux and discarded. In addition, images containing transits of Mars' moon across the Sun [Bell *et al.*, 2005] were flagged and discarded. One step that we omitted from the standard calibration is the flat field correction, which is discussed in the next subsection.

Determining flux from thumbnail images is conceptually similar, although only the image header is needed. On-board background subtraction removed contributions from dark current and detector bias, and integrated flux was returned in the header along with the location of the Sun in the original image. In the pipeline, saturation was flagged for a too-high integrated flux (determined by inspection of subframes with saturation), or a too-high maximum DN level in the thumbnail. Exposure time was then divided out. Such use of the header value is actually preferable to processing the downlinked thumbnail image due to the issue of lossy compression, as the 20-pixel Sun was downsampled generally into 1-4 pixels and then compressed from 12 to 1 bit per pixel. For thumbnail images flagged with possible saturation, only non-thumbnail images could be used, and only after verification that saturation did not occur.

The CCD response varies with wavelength and temperature [Bell *et al.* 2003]. In general, blue filters show decreased response for increasing temperature, and near infrared filters show the opposite, as was seen in the pre-flight calibration for L8 and R8, respectively. Following Bell *et al.* [2003] but rearranging the terms slightly, we model the response as linear with temperature, such that $F = F_{DNS} \cdot R_0 \cdot (1 + R_1 \cdot (T+15°C))$, where $F_{DNS}$ is the measured total DN/s, $R_0$ is the responsivity at -15°C, and $R_1$ is the responsivity slope, $T$ is the detector temperature in °C. A reference of -15°C was chosen as the average temperature for images taken in the first Mars year of the mission. We define $R_0 \equiv 1$ to maintain DN/s as the flux units of choice, but still account for the slope.

The temperature slope for the CCD using the R8 filters was reasonably well determined in preflight calibration, but that using the L8 filters was not; its calculation is described here. The preflight calibration values relied on comparison with 430- and 750-nm non-solar filters and an assumption regarding the red leak. Inspection of L8/R8 paired flux data shows that assumption to be wrong. If one assumes that $\tau_R = \tau_L \cdot (1+ \varepsilon)$—where $\varepsilon$ is a constant fractional difference in optical depth between the two filters—then the flux ratio for any given observation, $F_L/F_R$, is approximately linear in temperature with a slope of $R_{1R}-R_{1L}$ and approximately linear in $\tau_{slant}$ (the slant-path optical depth, equivalent to $\tau \cdot \eta$) with a slope of $\varepsilon$. We found that the average $R_1$ difference was $0.00411\pm0.00002$ and $0.00425\pm0.00004$ °C$^{-1}$ for Spirit and Opportunity respectively, while $\varepsilon$ averaged $0.0330\pm0.0004$ and $0.0256\pm0.0007$. The implication of the larger temperature sensitivity of the CCD with L8 compared to Bell *et al.* [2003] is that the ratio of in- to out-of-band flux in L8 may be 6:4 rather than 1:9, with the difference attributable to the relative blueness of the Sun compared to the calibration lamp and the fact that the in-band response was simply lost in the noise when imaging the calibration lamp. No trends or outliers were seen in the fit, suggesting that there were no detectable changes in $\varepsilon$ (for instance, due to the occasional presence of large ice particles). However, we calculate that $\varepsilon$ is only weakly size-dependent for dust with mean radius in the 1-5 μm range and a size distribution with variance 0.2, even without the red leak.

### 3.3. Flat field

Prior to summing pixels, a flat field image would normally be applied to account for high-frequency pixel-to-pixel variability and low-frequency variations across the detector caused by the optical system (intensity fall-off away from the optical axis, vignetting, out of focus dust on optical elements). However, accurate pre-flight solar-filter flat-field images could not be

obtained. Certain data sets (Sol B/260, P2694; B/290, P2694; B/1276, P2694; B/1588, P2679; A/312, P2699; A/1295, P2691) imaged the Sun repeatedly, changing only the position of the Sun on the CCD, allowing a test of the flat-field strategy. Against several options—nearest regular-filter flat-field, weighted-average of regular-filter flat-fields, a 2-dimensional polynomial fit and $\cos^4(\varphi)$ fall-off, where $\varphi$ is the angle off the optical axis—defining the flat field as unity produced the lowest $X^2$ for sols <1200. With respect to high-frequency variations of order <1%, this is not problematic, as 300 pixels were summed when determining the solar flux. Experience suggests that mid-frequency (one to several times the Sun's apparent diameter) variations could be of order 1%, and may be caused by out of focus dust in the optics and other sources. For Opportunity, after the dust accumulation event described below, a second order polynomial fit in line and sample was used to correct all observations for variations seen in P2694 and P2679. The corrections are up to 2% before the dust event and up to 23% after the dust event, increasing primarily to the right across the frame.

### 3.4. Changing window dust

The cycle of accumulation and removal of dust on the Pancam windows posed a problem in the determination of atmospheric opacity even for observations with the Sun in the same position. Interpretation of extinction from new window dust as new atmospheric dust was operationally significant, as the error would interfere with the assessment of the performance of the solar arrays. In addition, the retrieved window extinction came to exceed atmospheric extinction for parts of the mission.

Accumulation of dust was recognized prior to the sol ~1250 dust accumulation event. However, the accumulation event is illustrative. After the event—which occurred during the largest dust storm yet seen by the rovers—opacity was elevated. This was not diagnostic, as it could be considered routine when recovering from a planetary-scale dust cloud. However, diurnal variations—variations inversely correlated with airmass—were diagnostic. The process for identification and correction of dust events relies simply on a continuous reapplication of the Beer-Lambert Law. The transmission of atmospheric dust decreases with airmass, while the transmission of dust on the window is independent of airmass, and mimics a lower value of $F_{TOA}$. A 2-component model was used, where $F = F_{TOA} \cdot \exp(-\tau_{window} - \tau_{atmo} \cdot \eta)$; thus, $\ln(F) - \ln(F_{TOA}) = -\tau_{window} - \tau_{atmo} \cdot \eta$. Figure 4 shows the resulting fit for every sol with at least two Sun images taken over a range of >0.5 airmasses and at airmasses<5; error bars were calculated assuming the dominant error is natural variability in optical depth at the 5% level. A preliminary calibration was used to scale the data; for the final calibration, we asserted $\tau_{window} = 0$ at landing and adjusted $F_{TOA}$. That choice has no impact on the atmospheric optical depth retrieval. While no significant degradation of the CCDs over the mission is expected, it would be part of the correction attributed to window dust.

  The history of dust on the window can be inferred from Fig. 4; so too can limitations on the method. High-frequency variations are typically within the noise. Although we associate the slower variations with dust build up, the higher-frequency variations could result from an error in the temperature calibration modulated by shifting patterns in time-of-sol sampling. We chose to conservatively interpret the data into a dust history, by modeling dust build up as a series of linear changes, with the slope changing at discrete times. For the model, the regimes were chosen initially by hand; then within each time range, a linear fit for window-dust as a function of time was determined. The fits were spliced together, with their intersections defining the new regimes. Other functional forms, such as exponential decay, could be used, but we found none that provided a decrease in the residuals.

All accumulation-regime boundaries were assumed to be continuous in dust amount, but not in rate of accumulation. They are well-determined at the ~2% level, except for that near sol B/1260. For that change, fits are unavailable for sols 1216-1280. Individual Sun images are available for most sols, with some gaps such as 1236-1242 at the height of the dust storm when the rover had insufficient power. Inspection of solar array data shows rapid dust accumulation over sols 1254-1268. If the Pancam dust accumulation occurred at that time, then the pre- and post-peak comparisons of opacity and array data are in reasonable agreement. The retrieval was done using this assumption, but using an artificially large systematic error bar through this period The peak of the storm as seen in solar array data was sol 1237, when we estimate opacity at 4.8 from the array data, but when imaging was not commanded.

### 3.5. Airmass

Airmass is the ratio of path optical depth along an arbitrary line of sight to that in the zenith direction. Airmass is frequently approximated as $\sec(\theta)$, where $\theta$ is the solar zenith angle; this relation applies when the atmosphere thickness is small compared to the planetary radius. While we did not make that approximation, the distinction is immaterial for the vast majority of the data. We discuss here the minority of the data for which our assumptions regarding airmass matter. The extreme case is the moment of sunset: infinite airmass for a plane-parallel atmosphere, but airmass of about 22 for spherical shell atmosphere with a scale-height of 11 km and a radius of 3396 km.

When the Sun is below an elevation angle of 30° (with respect to the gravitational normal vector), the airmass begins to be sensitive to the vertical distribution of the dust. Above that elevation, and the airmass for a plane-parallel atmosphere (*e.g.*, all dust confined to a thin boundary layer such that the vertical scale is small compared to the planet's radius) and that for an 11-km scale height differ by <1%. Below 10° elevation, the difference exceeds 10% and the airmass becomes somewhat sensitive to the specific choice of scale height. Other measures of vertical structure are possible: while these data are sensitive to vertical structure, they are not diagnostic.

We determined an airmass relation to use for each rover by (1) finding all sols with valid data both with the Sun below 10° and above 30° elevation; (2) averaging all high-Sun (elevation>30°) data for each sol to get a characteristic mean opacity; (3) assuming the actual opacity when the Sun is low to be within ±5% (1-$\sigma$) of the mean; (4) taking the ratio of each observed slant opacity to the mean zenith opacity to be the airmass; (5) finding a single scale-height that best fits the observed distribution of airmasses. We chose a scale-height model in order to require only one new free parameter for vertical variation, due to the lack of uniqueness inherent in using airmass to infer vertical structure. The best fit scale-heights are 11.5±0.4 and 12.2±0.4 km for Spirit's L8 and R8, respectively; and 19.3±0.5 and 19.1±0.5 km for Opportunity's L8 and R8. For defining the elevation-airmass relation, we adopted model scale heights of 11.9±0.4 and 19.2±0.5 km, for Spirit and Opportunity respectively. We did not reduce the standard error values to account for averaging, as the sampling biases are similar for each filter. Implications are discussed in section 4. The derived scale heights for Spirit are similar to typical gas scale heights near the surface. Those for Opportunity are surprisingly high, and likely arise from a real distribution that is different from a purely exponential model, as discussed in section 4.5.

### 3.6. Precision and accuracy

Each individual measurement is a product of both a specific photon counting event and a relative calibration process. For nearly all data, counting statistics are irrelevant and systematic sources dominate the noise. The key relation can be written:

$$\tau_{slant} = \tau \cdot \eta = \ln(F_{TOA}) - \ln(F_{DNS}) - \ln(1+R_1 \cdot (T+15°C)) - \ln(T_{flat}) - \tau_{window}.$$

Therefore, when considering slant path opacity with a temperature correction <<1,

$$\sigma_{slant} \sim [\,(\sigma_{TOA}/F_{TOA})^2 + (\sigma_{DNS}/F_{DNS})^2 + (\sigma_{R1} \cdot [T+15°C])^2 + (\sigma_{flat}/T_{flat})^2 + \sigma_{window}^2\,]^{1/2}.$$

And in general for the reported normal-incidence opacity,

$$\sigma_\tau = \tau\,[\,(\sigma_{slant}/\tau_{slant})^2 + (\sigma_\eta/\eta)^2\,]^{1/2}.$$

For most data, $\sigma_\tau \sim \sigma_{slant}/\eta$, as the airmass uncertainty is negligible. For the other terms, counting noise was typically small: the ratio of noise to signal is approximately $N^{-1/2}$, where $N$ is the total number of photons in the exposure, typically $6 \times 10^5$. The read noise and uncertainty in the background subtraction were larger for low-signal images, and were explicitly considered in the error propagation. Uncertainty in the temperature term is a few times $10^{-4}\,°C^{-1}\,[T+15°C]$, which, if all data were taken at -15°C, would have no contribution. For data taken at CCD temperatures above or below the mean, this term contributes typical uncertainties of $3\text{-}5 \times 10^{-3}$. For the flat field correction, $\sigma_{flat}$ was assumed to be 0.02, with $T_{flat}$ typically near unity, and this was initially the dominant instrumental source of uncertainty.

Uncertainty due to non-photometric conditions (*i.e.*, non-uniform atmospheric dust) was the most difficult to quantify. We grouped the uncertainty associated with the top-of-atmosphere fit and the window dust model, as they have identical impact on a Beer-Lambert fit at any given time during the mission. Figure 4 shows error bars in individual fits, as determined from the uncertainties noted above and an assumption that natural dust-loading variability is 5%, which was determined to be consistent with the quality of the individual fits. Note that we omitted from these fits any sols during times of rapid sol-to-sol changes such as local dust storms or regional dust storm onset. While treating each measurement as independent would result in a lower estimate for the uncertainty in the fit, we took the largest systematic departures from the fit to imply the error associated with the top-of-atmosphere flux and window dust to be 0.03 (1-$\sigma$) for each filter, except 0.04 for Opportunity's R8. In addition, we increased the uncertainty for times of rapid window-dust accumulation by the net change in 10 sols; and we increase the uncertainty for sols A/1500-1800 and B/1216-1280 by 25% of the modeled window dust. These latter changes represent our estimate of a prudent systematic error to account for times when the model is poorly constrained.

The final source of uncertainty to note is that associated with airmass. Two factors are the uncertainty related to choice of model and the uncertainty relating to time. For the former, we assumed the scale-height model is an adequate description of the airmass variation with solar elevation, and we used the nominal scale-height uncertainties. For the latter, we adopted a time uncertainty of ±15 s based on the project requirement to maintain time knowledge (the translation from the spacecraft clock to spacecraft event time in UTC, from which the Sun's position is calculated) of 30 s in the worst case. This translates to an elevation error of <0.1°. Airmass errors are irrelevant when the Sun is above 15°, but dominate immediately before sunset when they approach 10%.

The resulting error represents systematic and random effects as discussed above. In most cases, an absolute error of ~0.03/$\eta$ (~0.04/$\eta$ for Opportunity's R8) can be assumed to be systematic—resisting attempts to use averaging to reduce error—while the (quadratic) remainder can be assumed to be random. Errors in high-airmass (>10) optical depths are dominated by systematic effects. Averaging to reduce the error is not recommended due to the identified systematic terms. A conservative approach would be to make comparisons across and within sols with consideration for their (quadratic) cumulative error.

### 3.7. Planetary Data System archive

The data are archived with the Planetary Data System [Lemmon, M., MER Mars Pancam Atmospheric Opacity RDR V1.0, NASA Planetary Data System, MER-M-PANCAM-5-ATMOS-OPACITY-V1.0, 2004]. There are two files for each combination of rover and camera, a label (*.LBL) and tabulated data (*.TAB). An example filename is "2TAU_440_2209_20110218A.TAB", where "2" signifies MER-A, Spirit ("1" is MER-B, Opportunity); "440" is the nominal wavelength of the L8 filter ("880" is the value for R8); "2209" is the last sol appearing in the table; "20110218" is a string signifying the table creation date (YYYYMMDD). As is customary, the labels describe the contents of the tables. The tables themselves include columns for EDR product identifier, $L_S$, Mars-Sun distance, decimal sols since midnight before landing (sol number-1+LTST/24—one rounds up to get project sol number), airmass, flux in corrected DN/ms, optical depth, and 1-σ error in optical depth.

The results discussed here refer to the calibration associated with the time stamps "20110218" for Spirit and "20131028" for Opportunity. Future deliveries will be updated according to continued application of the techniques described here.

## 4. Discussion

### 4.1. Seasonal variations of optical depth

Figure 1 shows the optical depth record of the mission. Values in both filters are nearly the same: values discussed in this section are from R8 (880-nm) unless specified otherwise. Due to the well-understood spectral sampling, the R8 opacity is recommended as the value of reference. The rovers' initial 90-sol mission included the southern autumnal equinox ($L_S=0°$) that began MY 27, during a time of moderate but declining optical depths [Lemmon *et al.*, 2004] following a regional dust storm that was dissipating by the landing time [Smith, 2008]. The TES climatology [Smith, 2004] would suggest that optical depths would decline into aphelion season for both rovers, rise slightly near $L_S=135°$, and become dusty during perihelion season, with dust increases around $L_S=180°$, $L_S=220°$, $L_S=320°$. Spirit optical depths declined from near 0.90 to below ~0.3 by sol 155 ($L_S=45°$), and remained similarly low until about sol 350 ($L_S=135°$). During this time, Opportunity optical depths declined from 0.95 to below ~0.5. The sudden increase of optical depth over $L_S=135-137°$ was an operational surprise. It occurred in the last week of 2004, during a time of reduced staffing and was seen as the first major reduction in solar power of the mission: an optical depth 1.24 dust storm peaking over sols B/329-332. Both rovers experienced local dust storms with optical depths of 1-1.5 and generally elevated optical depths from $L_S=135°$ to 200° (to about sol 500 for Spirit).

For context, Mars Global Surveyor (MGS) Mars Orbiter Camera (MOC) views of some of these storms are shown in Fig. 5. The MGS spacecraft reached its designated 380-km altitude, 1400 LMST equator crossing (south-to-north), circular polar orbit in March 1999 and operated through November 2006. For that time, MOC provided nearly continuous daily global mapping of the Martian surface at 7.5 km pixel$^{-1}$ in two band passes, red (600-650 nm) and blue (400-450 nm) [Malin *et al.* 2010]. In the spirit of mutual collaboration between missions, MOC and later the Mars Reconnaissance Orbiter (MRO) Mars Color Imager (MARCI) provided almost daily regional weather updates and forecasts for the two MER sites during dusty seasons.

From late December 2004 ($L_S=131°$) through March 2005 ($L_S=185°$) there were several significant dust events, which either passed through or near the Meridiani and Gusev Crater rover sites. The earliest was the most extensive of these dust events, observed as the MGS spacecraft came out of contingency mode on 27 December 2004, $L_S=135°$ [Cantor, 2007]. Prior to that time, weather conditions at southern low-latitudes were normal for mid-winter. The aphelion water-ice cloud belt was at its maximum extent, orographic water-ice clouds were observed over the Tharsis volcanoes, and an occasional local dust storm in noted Syria. By $L_S=135.5°$, dust storm activity extended from 0°-40°S, 300°-90°W with most of the activity occurring in Noachis and Valles Marineris, as well as locally in Margaritifer. Storm activity extended into Meridiani by $L_S=137.0°$; see Fig. 5a. The origin of the dust event was not known exactly, due to the absence of MOC observations during the onset of storm activity. But it was clear from the first observations, that the dust event was not a single storm but consisted of at least one regional storm in Noachis with local storms in Valles Marineris, Margaritifier, and Meridiani. The storm activity was extensive enough to generate a diffuse dust haze that encircled the southern hemisphere within a two week period, raising the visible background opacity above nominal levels for this time of year, to 0.7 at both sites. Cantor [2007] noted that this event was seasonally the earliest planet–encircling dust veil observed to date. Dust-lifting activity abated by $L_S=138.5°$, only several sols after it was first observed. However, the background opacity in the southern hemisphere remained elevated for over a month, which resulted in water ice clouds associated with the aphelion cloud belt and the shield volcanoes to disappear for that period of time.

Within the 70 sols after the planet-encircling dust veil, a series of dust storms passed over the Spirit site. The first was observed by MGS on the 19th of January 2005 ($L_S=146.5°$); see Fig 5b. Initially the local storm formed in Promethei along the south polar cap edge. The storm moved east-northeast at ~13 m s$^{-1}$ over a 2-sol period before crossing over Gusev crater. The storm quickly abated the following sol, but a diffuse dust haze remained over the site for several additional sols. The next storm fronts to reach Gusev were observed from 6-10 March ($L_S$=170.6-173.0°, sols 417-421). That activity consisted of two storms, a single regional storm that passed through the Spirit site on 7 March and a large local storm that came through two sols later. Both storms began along the edge of the seasonal receding south polar cap edge. The regional storm began as a large local storm in Promethei on $L_S$=168° and over a 2-sol period moved east at an average speed of about 22 m s$^{-1}$, merging with local storms along the cap edge and dramatically increasing in area. The storm reached its maximum areal extent (~6.4x10$^6$ km$^2$) at $L_S$=169°. At that same time, the leading edge of the storm was in Sirenum, far to the southeast, while the western trailing edge extended south-southwest of Gusev. The western, trailing edge of the storm moved north over the Spirit site at $L_S$=171°. The storm was gone as quickly as it came, moving off to the east by the following sol. The second storm, developed just to the south of Gusev in Cimmeria. The storm moved northward at about ~13.5 m s$^{-1}$, completely enveloping the Spirit site at $L_S$=173°; see Fig. 5c. The storm, like the previous local storms, had dissipated by the following sol, but it left behind a diffuse dust haze that persisted for several additional sols.

Near $L_S$=215-220°, a new period of rising opacity was seen first at the Spirit site, but briefly peaked at optical depth 1.98 at the Opportunity site on B/492, as cross-equatorial storms on the Acidalia storm track created a second planet-encircling haze [Cantor, 2007]. Optical depths fell gradually until $L_S$=307° before the next wave of cross-equatorial storms increased opacity to 1.76 on sol B/630 ($L_S$=316.5°) and eventually encircled the planet. A Hubble Space Telescope (HST) Advanced Camera for Surveys (ACS [Sirianni *et al.* 2005]) image (Fig. 6) shows a 1500-km long regional dust storm in Meridiani on sol 626 ($L_S$=314.2°, 28 October 2005) just before the storm peaked at the Opportunity site. After this time, optical depths at both sites gradually declined while the mission entered its second Mars year. By $L_S=0°$, the observations essentially matched those of the initial 90-sol mission.

Subsequent Mars years displayed broadly similar seasonal behavior, with the major exception being the globally extensive dust storms around $L_S=270°$ of MY 28. Figure 7 shows optical depths for all years in their seasonal context. The optical depths generally match orbital climatology from imaging and infrared opacity, which show low optical depths through southern winter, receding cap-edge dust storms leading to an increase in dust after $L_S=135°$, and regional storms around $L_S=220°$ and $315°$ [Smith 2004, 2009; Wang 2007].

The period from $L_S=0-135°$ was substantially the same in each year. Optical depths at the Spirit site show a fall-off over $L_S=10-50°$, followed by a slight rise through $L_S=70$, a decline to $L_S=85°$, a rise to $L_S=105°$, and a decline to $L_S=135°$. The features are low-amplitude, but repeatable, and suggest an influx of dust around 15-20° before and after northern summer solstice. The most notable departures occur during MY 29—few calibration observations were able to be acquired during this time, and the systematic uncertainties are at their largest. It is reasonable, but not necessary, to suppose that the calibration is off and MY 29 should be offset slightly lower to mimic the other years. At the Opportunity site, the first year's pattern repeats, including an inflection in the slope of decline at $L_S=75°$. During this time, several individual data points lie above the trend as seen with both filters. These are likely due to discrete clouds, as discussed in section 4.3.

The period from $L_S=135-215°$ was similar to the first year. The optical depths did not repeat as they did in the winter, but the period was dominated by generally increased opacity. The period did not start with as sudden a rise as in the first year, but the rovers typically experienced ~1 small storm of a few sols in duration and peak optical depths of over 1 to about

2. MY28 is notable in showing a more gradual increase in Meridiani than the other Mars years. The dust load away from local storms tended to vary by tens of percent across years, and the timing and number of local storms (or brief opacity spikes) varied.

For $L_S$=215-315°, MY 29 is almost identical to MY 27, with regional dust storms increasing opacity at both sites early in the period and then a general decline. MY 30 and 31 are similar at the Opportunity site, with a later onset of diffuse opacity from regional storms and smaller opacity increases. For years other than MY 28, the optical depths from $L_S$=240-305° are remarkably consistent, especially for Opportunity (Fig. 8).

MY 28 is distinguished by a somewhat steady and relatively low optical depth around $L_S$=215-265°, followed by a rapid increase beginning at $L_S$=265° for Opportunity and 269° for Spirit (26 June and 2 July 2007). This storm led to optical depths up to 4.3 at the Spirit site, where the optical depth was about 3 or more for a 30-sol period and rover operations were dramatically slowed by lack of available solar power. At the Opportunity site, the observed peak optical depth of 4.6 occurred on sol 1235 ($L_S$=277, 16 July 2007), after which the rover was commanded to hibernate to slow the drain of the batteries. The next observations, on sols 1243 and 1246, showed optical depths down to 3.9 and then back up, to eventually reach 4.5 and then slowly decline. Analysis of solar array data suggests the peak sol-averaged optical depth was probably near 5, comparable to estimates of the first VL1 storm, but short of the second, $\tau$=9 storm [Pollack *et al.,* 1979]. MRO MARCI [Bell *et al.* 2009] images were used to track this event globally. Figure 9 shows the Opportunity site as a series of individual storms pass near the site and the general dustiness of the atmosphere increases. After $L_S$=295°, both sites show a similar, steady decrease of opacity (see Fig. 8). The initial rate of decrease is 2.3-2.4%/sol, before the decrease slows and the optical depths become similar to those from other years.

Smaller dust storms typically had higher peaks rates of decline maintained for shorter durations, indicating advection as a likely means of removal. Global dust storms reported by Viking initiated near $L_S=205°$ and reached optical depth 2.7-3.2 at the VL1 site; and $L_S=270°$, reaching optical depth 3.7-9 at the VL1 site [Pollack *et al.,* 1979].Our decay timescales of 43 sols for each rover and then 77 sols for Opportunity are similar to the decay rates following the VL1 second global storm of 51 and then 119 sols [Pollack *et al*., 1979]. The lower rate, which was observed for both VL1 storms but not for Spirit, is likely due to renewed dust lifting in the Acidalia cross-equatorial storm track.

As observed from orbiters, the period from $L_S=315-360°$ typically produces a final wave of cross-equatorial dust storms [Cantor *et al.,* 2001;Wang *et al.,* 2002, 2007; Cantor, 2007], resulting in elevated dust optical depths followed by settling at the rover sites. This activity, as seen at the rover sites, was sporadic, with decay of the global event at both sites in MY 28, regional storms at the Spirit site in MY 29 and the Opportunity site in MY 27, 29, and 31 (and prior to landing), and general lifting otherwise.

The overall picture is qualitatively the same as previous reports. Smith [2004] presents annual and zonal averaged TES-derived optical depths at 9 μm for MY 24-26. The seasonal trends are the same as those seen by the rovers. A gradual decline in dust optical depth to $L_S=140°$ is followed by elevated backgrounds, then further enhanced storms (and the 2007 global dust event) after around $L_S=220°$, and then a third wave of lifting near $L_S=340°$. Smith [2009] presents a similar data set for MY 26-29 from THEMIS, substantially overlapping the rover mission. These show the same pattern. In both cases, the 9-μm data are about half the Spirit observations at the same latitude. The differences between the data sets come mostly from the visible vs. infrared properties of the dust and the presence of water ice at the Meridiani site in the winter (both discussed below) and slightly from meridional variations.

Away from the globally extensive dust event, the opacity declines around $L_S$=0, 180, and 270° occur when lifting and transport may supply or remove some dust. We considered every 10-sol period in the mission in 5-sol increments, measured the goodness of fit of an exponential change model and the associated rate of change. To investigate periods of steady decline, we exclude periods for which the data were poorly fit (such as the passage of local storms) and periods that had dust increasing or stable (slope consistent with 0 or less at the 90% confidence level). We found remarkable stability for each rover outside of the storms mentioned above. For Opportunity, the average change during periods of steady decline was 0.52±0.05% per sol; for Spirit the decline was 0.77±0.06% per sol. The rates and their implied removal timescales of 130-200 sols are consistent with initial reports of 0.6-0.7% per sol by Lemmon *et al.* [2004] for the spring period. The rate is also consistent with the settling rate determined for Pathfinder of 0.28%/sol by surface coverage [Landis and Jenkins, 2000] during a period of optical depth ~0.5 [Smith and Lemmon, 1999], implying an atmospheric loss to sedimentation of 0.56%/sol. Thus sedimentation is likely a significant component of dust loss through much of the year. Given the timescales above, areas that are net sinks for atmospheric dust could gain up to 3-5 times the mean atmospheric dust load each Mars year.

**4.2. Seasonal variations in aerosols**

In principle, varying color ratios can be indicative of aerosol size changes, and color ratio (or Angstrom exponent) has some diagnostic value. We find that (1) the ratio of the L8 to R8 opacity does not show significant variation during the first part of the mission; and (2) variability late in the mission is unmeasurable due to uncertainties in the correction of variable dust on the windows. Section 3.2 reports the mean ratio of (R8-L8)/R8 opacity as 0.0330±0.0004 and 0.0256±0.0007. With the nominal factor of two wavelength difference, the ratios correspond to an Angstrom exponent near zero, typical of large particles. Dust aerosol size distributions have been reported with sizes around 1.5-1.65 µm (radius of equivalent volume sphere) and variances around 0.2-0.5 for Viking, Pathfinder and MER [Pollack *et al.* 1995; Tomasko *et al.* 1999; Lemmon *et al.* 2004]. Such aerosols have an extinction maximum near or longward of 1000 nm, and have about 7-11% less opacity at 440 nm than at 880 nm. Given the red leak for the L8 filter, we cannot exclude such aerosols.

For the first part of the mission we have measurements of 9-µm dust opacity from Mini Thermal Emission Spectrometer (Mini-TES) data, as reported by Smith *et al.* [2004, 2006]. The record was continued through the middle of MY 28 before dust on the Mini-TES optics rendered the data useless. Figure 10 shows the Mini-TES data with the 880-nm data for both rovers—note the infrared opacity scale is half the visible scale to generally overlay the data. All data are shown except three measurements near optical depth two on sol A/1243. These were two times the 880-nm measurements at the time, but are subject to increased uncertainty due to the dusty optics. The infrared/880-nm ratio is shown in Fig. 10(C). Convention is to report the inverse as a "visible/IR" ratio. However, due to the distribution of error bars, the figure is more interpretable showing "IR/visible". There are clear trends: after landing, the ratio falls for both rovers; Opportunity shows more complex behavior during the first northern summer; the ratio increases significantly with the onset of dust storms; the ratio tends to decline after dust storms and generally increases again for new storms; the ratio then falls for the next northern summer and rises with the onset of the second dusty season. Interestingly, the $L_S$=220° dust storm did not change the ratio at the Spirit site. Note that the ratio of sol-average optical depths is shown, so the few sols during which there was notable variability may have incorrect ratios.

The comparison of these optical depths can be informative as to dust aerosol size [Clancy *et al.* 2003; Wolff *et al.* 2006]. For certain assumptions about IR optical properties of dust, Wolff *et al.* [2006] derived the (inverse) ratio as a function of mean size and variance of the size distribution. Based on their Fig. 13 and assuming a variance of 0.5, in Fig. 10(C) we show the ratio for mean radii of 0.7, 1.4, and 2.1 μm for both rovers. For different assumptions, those sizes would change, but the approximate relation would remain. The Opportunity data around $L_S$=30-130° are somewhat problematic: the 880-nm value is a total optical depth; the Mini-TES value is specific to dust. Mini-TES did not detect water ice, but orbital data show ice at that time, and the Mini-TES data do not rule ice out at high altitude [Smith *et al.* 2006]. So, the ratio may be depressed due to ice hazes. It is possible that the increase in the Mini-TES /880-nm ratio during this period may indicate a sensitivity to ice-coated dust. Generally, dust sizes fall as low as ~1 μm in northern summer and increase to over 2 μm with the onset of a dust storm (given the caveat above). This pattern is similar to patterns reported by Clancy *et al.* [2003] using orbital infrared spectroscopy, and generally consistent with dust sizes of 1.2-1.4 μm during the northern summer at the northerly Phoenix site [Komguen *et al.*, 2013].

### 4.3. Clouds

The Spirit and Opportunity optical depth records are quite similar. One systematic exception is the elevated opacity during northern summer at the more northerly latitude. Additionally, during the $L_S$=0-135° time, Opportunity optical depths tend to steadily decline, with infrequent outliers (in both filters) of magnitude ~0.1 above the trend. While Mini-TES did not directly detect ice aerosols, possibly due to their location at cold, high altitudes, the TES climatology and contemporaneous orbital observations do show near-equatorial ice aerosols [Smith 2004, 2009]. The TES climatology shows slightly more dust at the Opportunity site compared to the Spirit site, but also shows much more ice, during the $L_S$=0-150° timeframe, with a sudden decrease at $L_S$=135° during the first year. For each rover, this is the only time of year when appreciable ice-aerosol amounts are expected. While separation of ice and dust was a motivation for inclusion of Pancam's two solar filters [Bell *et al*. 2003], the sensitivity is only to particles of very different sizes, and is significantly reduced due to the L8 red leak.

During the first two years (after sol 150) regular sky monitoring was done with the rovers' Navcams, taking advantage of the 45x45° field of view [Maki *et al*. 2003]. Throughout the missions, regular drive-direction and other panoramic imagery frequently showed the sky. Under typical aphelion-season conditions, we expect line-of-sight opacity variations in excess of a few percent to be detectable, although the limit does increase for dustier skies. Prior to sol B/100, no discrete clouds were noted. During sols B/101-109 ($L_S$~30°) and again in the 140s ($L_S$~50°), clouds were noted in several images, leading to the regular campaigns (Fig. 11). During the first Mars year, cirriform clouds were seen occasionally until sol B/310 (over $L_S$=50-126°), especially around $L_S$~55° and 115°. During the second year, cirriform clouds were seen over $L_S$=23-50° and $L_S$=110-131°. During the third year, sampling was more sparse, but such clouds were seen near $L_S$=20° and over $L_S$=109-136°, with a real gap in between. Additional clouds were seen near $L_S$=30° and $L_S$=63° of the next two years (successively), but sampling had been curtailed due to operational constraints. Such clouds are likely the reason for the small spikes in optical depth. Obviously-cirriform clouds were not observed at the Spirit site, although some of the infrequent wave structures were likely due to condensates.

The images are not diagnostic as to the cloud composition. Water ice is likely, given the aphelion cloud belt over the Opportunity site. However, mesospheric carbon dioxide ice clouds have been seen near the Meridiani area, peaking near $L_S$=30 and 150° [Clancy *et al.*, 2007]. It is possible that some of the early-season clouds were $CO_2$ ice. Figure 11 shows some of the variety of clouds, including wave and more complicated structures, not unlike terrestrial cirrus.

The cloud detections were episodic, with many over a few sols and then many sols in between them. Yet the elevated opacity was consistent. Thus, the general opacity at the Opportunity site during northern summer was either dust, or dust with a somewhat uniform ice haze that was not detectable as discrete clouds. No discrete ice clouds were seen at either site outside of $L_S$=20-136°.

**4.4. Seasonal variations in insolation and dust devils**

Dust devils are seen across Mars in orbital imagery [Malin and Edgett, 2001; Cantor *et al.*, 2006], and they have been reported and used to estimate local dust lifting rates at the Mars Pathfinder, Spirit, and Phoenix landing sites [Metzger *et al.*, 1999; Ferri *et al.*, 2003; Greeley *et al.*, 2006a, 2006b, 2010; Ellehoj *et al.*, 2010]. We note that in Opportunity images, dust devils do appear and dust is seen being lifted by wind gusts in craters; however, such occurrences are rare compared to the Spirit site, presumably due to the lower availability of surface dust and the generally smoother surface.

Heating of the surface and atmosphere by sunlight is governed by both orbital parameters and atmospheric extinction. Mars has an eccentric orbit, with aphelion at 1.67 AU and perihelion at 1.38 AU. Thus, globally averaged insolation at the top-of-atmosphere (TOA) varies by nearly 50% between these two orbital positions. The Opportunity site is nearly equatorial, so seasonal effects are dominated by the Sun being lower in the sky near the solstices. The Spirit site is still

"tropical" in the sense of having the sub-solar latitude sweep over its latitude before and after summer solstice. However, it is sufficiently far south that the Sun is much lower in the sky in winter. The two effects are roughly in phase, with perihelion at $L_S=251°$, roughly 35 sols before summer solstice.

We have modeled the solar fluxes at TOA, at the surface and absorbed into the atmosphere (Fig. 12). The TOA calculations are straightforward geometric calculations, taking into account the varying distance to the Sun, length of daytime, and elevation of the Sun throughout the day at each site. Atmospheric calculations were done with the Tomasko *et al*. [1999] dust and radiative transfer model, supplemented by expanded dust parameters provided Johnson *et al*. [2003], every 15 minutes for the duration of the mission. The sol-averaged optical depth value was used in each calculation. Wavelengths outside of 440-1000 nm were extrapolated from the above parameters and contributed a minority of the flux. Atmospheric gas absorptions were ignored, and we consider only solar radiation, rather than thermal radiation. These models are similar to project-internal models used to monitor the solar panels and to estimate power loss due to dust deposition. One distinction is that this paper does not consider the effect of rover tilt and instead focuses on insolation over a level surface.

Surface insolation is shown as flux incident on the surface, roughly 80% of which is absorbed at the Spirit site, 86% at the Opportunity site [Bell *et al.,* 2008]. Insolation follows the seasonal TOA trends, but is modulated by dust in the atmosphere. Even during the optical depth >4 dust storm, surface insolation only falls by a factor of ~2-3 (rather than ~100 for direct sunlight) due to scattered light reaching the surface. The figure shows total atmospheric column absorption, which is roughly correlated with optical depth. When light interacts with dust, the most likely outcome is forward scattering, so the proportionality of absorption to abundance holds to optical depths ~2; at higher optical depths, absorption increases less rapidly. For low optical depths, similar patterns would be seen at all levels of the atmosphere. For the highest optical depths, absorption is shifted higher in the atmosphere; however, this is not an especially strong effect due to the forward scattering. High levels of dust could be expected to play a stronger role in cooling the bottom of the atmosphere in the infrared, through increased radiative coupling to a cooled surface and to space. Note that $L_S=135°$ dust storms can reduce the solar flux to mid-winter levels at each site. The global dust event reduced flux at the Opportunity site to levels far below those in winter; Spirit did not see such a low value due to the near-solstice timing and a more southerly latitude.

Greeley *et al.* [2010] discuss 3 Martian "dust devil seasons" that occurred during the 3 southern summers of the Spirit mission. Figure 12(A) shows their estimates of dust devil density. We have added error bars and upper limits based only on the counting statistics reported in Greeley *at al.* [2010, Fig. 3], with no adjustment for time of day effects or any bias resulting from the detection method. We note that at the site, and for dust devils visible to the rover's Navcam during the warm time of day sampled in the Greeley *et al.* data, dust devil number density seems to increase exponentially with surface insolation (a similar fit could have been achieved for noon insolation or for the surface-atmosphere difference) such that the dust devil density is $10^{-3}$ km$^{-2}$ at 116 W m$^{-2}$ with an *e*-folding scale of 18 W m$^{-2}$. Other forms, such as a linear fit with a threshold near 140 W m$^{-2}$, are not ruled out due to the scarcity of reported data away from "dust devil season". The exponential relationship between dust devil number density and mean surface insolation fits the seasonal change, the effect of the MY 28 dust storms, and the fall off of dust devils with the pre-solstice ($L_S$~220°) dust storms of MY 27 and 29 and the final major storm of the mission ($L_S$~320°). Unlike a threshold model, the exponential is consistent with the formation of a new dust devil track near the rover at $L_S=17\pm5°$ during the first Mars year, far from "dust devil season" [M. C. Malin, *et al.*, Wheel tracks from landing site to hills, NASA's Planetary Photojournal (http://photojournal.jpl.nasa.gov/), PIA07192, 3 January 2005]. The specific relationship cannot be extrapolated across Mars (or even to the Opportunity site), as dust supply and local boundary layer meteorology are also integral factors in dust devil formation.

### 4.5. Vertical structure

To process data taken with the Sun at a low elevation, we invoked an atmospheric model with exponential fall-off of dust with altitude, such that the vertical distribution is fully expressed as a scale height in section 3.5. This convenient 1-parameter model is adequate with respect to the data, but it provides limited constraints on the vertical behavior of the dust. More complex dust structures [Whiteway *et al*., 2009; Heavens *et al*., 2011] are not ruled out. The Spirit data are characterized by a scale height comparable to what one would estimate for a Martian $CO_2$ atmosphere with a mean temperature of 210 K. The derived Opportunity scale height is significantly larger than the gas scale height. Such a scale height is unlikely, as it would imply a dust source high in the atmosphere and sink at the surface. However, any model that fit the data would share the characteristic that dust is preferentially farther away from the surface than would be expected for a well-mixed atmosphere. For example, a possible alternative is that, while the Gusev atmosphere has dust well-mixed for the bottom 2-3 scale heights (the region the model is sensitive to), the Meridiani atmosphere is top-heavy, perhaps reflecting the high-altitude tropical dust maximum at 20-30 km [Heavens *et al.* 2011]. Alternatively, a relative depletion of dust in the boundary layer is consistent with the relatively lower dustiness of the Opportunity site relative to the Spirit site. In any case, while one can conclude the dust is relatively higher at the Opportunity site, the form of model applied is not uniquely constrained.

## 5. Conclusions

1. A precise optical depth record for the Mars Exploration Rover sites has been created, using direct solar extinction measurements, and archived. The data set spans more than 3 Mars years for Spirit and more than 5 Mars years for Opportunity, including periods of local, regional and globally extensive dust storms. The data set shows that, for each site, the $L_S$=0-135° period is consistent year-to-year, with relatively lower dust amount; , the $L_S$=135-215° period has variable local to regional storms; the $L_S$=215-360° includes a global dust storm, but otherwise shows a similar set of dust lifted by regional storms. The 880-nm (Pancam R8 filter) optical depth should be used for comparisons among data sets and model inputs, as the L8 (440-nm) filter has a significant 750-nm contribution.

2. Comparison of the 880-nm data to contemporaneous dust optical depth measurements from Mini-TES shows that changes in the Mini-TES/880-nm ratio frequently occur with the onset and decay of dust storms. The values are consistent with variations in the mean radius of dust aerosols from near 1 μm during northern summer to over 2 μm at the onset of a storm, and being generally near the reported values of 1.4 μm (from observations near the beginning of the mission).

3. Ice clouds contributed to the observed opacity at the Opportunity site. Water ice hazes may have been present. Clouds were seen over $L_S$=20-136°, with peak activity near $L_S$=50 and 115°. Ice clouds and probably hazes were not a significant part of the opacity at the Spirit site.

4. The shortwave optical depth of the atmosphere drives important changes to surface insolation and to atmospheric heat input. Dust devils are common at the Spirit site when insolation is high, but are rarer when insolation is seasonally low or low due to dust storms. While the occurrence of dust devils in an area responds to many factors, the relative abundance across seasons correlates with surface insolation.

5. Data from the Spirit site are consistent with the dust being well mixed with the atmosphere in the bottom 10-20 km. The Opportunity data require relatively more dust at higher altitudes, perhaps reflecting the high-altitude tropical dust maximum at 20-30 km.

## Acknowledgements

We thank the entire MER science and operations team, whose daily efforts have led to the existence of this data set. We thank David Kass for a helpful review of the manuscript. This work was funded by NASA through the Mars Exploration Rover Project, a portion of which was carried out at the Jet Propulsion Laboratory, California Institute of Technology, under a contract with the National Aeronautics and Space Administration.

## References

Arvidson, R. E., S. W. Squyres, R. C. Anderson, J. F. Bell III, J. Brückner, N. A. Cabrol, W. M. Calvin, M. H. Carr, P. R. Christensen, B. C. Clark, L. Crumpler, D. J. Des Marais, C. d'Uston, T. Economou, J. Farmer, W. H. Farrand, W. Folkner, M. Golombek, S. Gorevan, J. A. Grant, R. Greeley, J. Grotzinger, E. Guinness, B. C. Hahn, L. Haskin, K. E. Herkenhoff, J. A. Hurowitz, S. Hviid, J. R. Johnson, G. Klingelhöfer, A. H. Knoll, G. Landis, C. Leff, M. Lemmon, R. Li, M. B. Madsen, M. C. Malin, S. M. McLennan, H. Y. McSween, D. W. Ming, J. Moersch, R. V. Morris, T. Parker, J. W. Rice Jr., L. Richter, R. Rieder, D. S. Rodionov, C. Schröder , M. Sims, M. Smith, P. Smith, L. A. Soderblom, R. Sullivan, S. D. Thompson, N. J.

Tosca, A. Wang, H. Wänke, J. Ward, T. Wdowiak, M. Wolff, and A. Yen (2006), Overview of the Spirit Mars Exploration Rover Mission to Gusev Crater: Landing site to Backstay Rock in the Columbia Hills, *J. Geophys. Res.*, **111**, E02S01, doi:10.1029/2005JE002499.

Bell, J. F., S.W. Squyres, K.E. Herkenhoff, J.N. Maki, H.M. Arneson, D. Brown, S.A. Collins, A. Dingizian, S.T. Elliot, E.C. Hagerott, A.G. Hayes, M.J. Johnson, J.R. Johnson, J. Joseph, K. Kinch, M.T. Lemmon, R.V. Morris, L Scherr, M. Schwochert, M.K. Shepard, G.H. Smith, J.N. Sohl-Dickstein, R.J. Sullivan, W.T. Sullivan, and M. Wadsworth (2003), Mars Exploration Rover Athena Panoramic Camera (Pancam) investigation, *J. Geophys. Res.*, **108**(E12), 8063, doi:10.1029/2003JE002070.

Bell, J.F., III, M.T. Lemmon, T.C. Duxbury, M.Y.H. Hubbard, M.J. Wolff, S.W. Squyres, L. Craig, and J.M. Ludwinsky, 2005. Solar eclipses of Phobos and Deimos observed from the surface of Mars. *Nature* **436**, 55-57, doi:10.1038/nature03437.

Bell, J. F., III, J. Joseph, J. N. Sohl-Dickstein, H. M. Arneson, M. J. Johnson, M. T. Lemmon, and D. Savransky (2006a), In-flight calibration and performance of the Mars Exploration Rover Panoramic Camera (Pancam) instruments, *J. Geophys. Res.*, **111**, E02S03, doi:10.1029/2005JE002444.

Bell, J. F., III, D. Savransky, and M. J. Wolff (2006b), Chromaticity of the Martian sky as observed by the Mars Exploration Rover Pancam instruments, *J. Geophys. Res.*, **111**, E12S05, doi:10.1029/2006JE002687.

Bell, J. F., III, M. S. Rice, J. R. Johnson, and T. M. Hare (2008), Surface albedo observations at Gusev Crater and Meridiani Planum, Mars, *J. Geophys. Res.*, **113**, E06S18, doi:10.1029/2007JE002976.

Bell, J. M., S. W. Bougher, and J. R. Murphy (2007), Vertical dust mixing and the interannual variations in the Mars thermosphere, *J. Geophys. Res.*, **112**, E12002, doi:10.1029/2006JE002856.

Bell, J. F.III, M. J. Wolff, M. C. Malin, W. M. Calvin, B. A. Cantor, M. A. Caplinger, R. T. Clancy, K. S. Edgett, L. J. Edwards, J. Fahle, F. Ghaemi, R. M. Haberle, A. Hale, P. B. James, S. W. Lee, T. McConnochie, E. Noe Dobrea, M. A. Ravine, D. Schaeffer, K. D. Supulver, P. C. Thomas (2009). Mars Reconnaissance Orbiter Mars Color Imager (MARCI): Instrument description, calibration, and performance. *J. Geophys. Res.,* **114**, E08S92, doi:10.1029/2008JE003315.

Cantor, B. A., P. B. James, M. Caplinger, and M. J. Wolff (2001), Martian dust storms: 1999 Mars Orbiter Camera observations, *J. Geophys. Res.*, *106*, 23653-26687.

Cantor, B. A. . K. K. Kanak, and K. S. Edgett (2006). Mars Orbiter Camera observations of Martian dust devils and their tracks (September 1997 to January 2006) and evaluation of theoretical vortex models, *J. Geophys. Res.*, **111**, E12002, doi: 10.1029/2006JE002700.

Cantor, B. A. (2007), MOC observations of the 2001 Mars planet-encircling dust storm, *Icarus*, **186** 60-96, doi: 10.1016/j.icarus.2006.08.019.

Clancy, R. T., M. J. Wolff, and P. R. Christensen (2003). Mars aerosol studies with the MGS TES emission phase function observations: Optical depths, particle sizes, and ice cloud types versus latitude and solar longitude. *J. Geophys. Res.*, **108**, E9, 5098, doi: 10.1029/2003JE002058.

Clancy, R. T., and S. W. Lee (1991), A new look at dust and clouds in the Mars atmosphere: Analysis of emission-phase-function sequences from global Viking IRTM observations, *Icarus*, **93**, 135–158, doi:10.1016/0019-1035(91)90169-T.

Clancy, R. T., B.J. Sandor, M.J. Wolff, P.R. Christensen, M.D. Smith, J.C. Pearl, B.J. Conrath, and R.J. Wilson (2000). An intercomparison of ground-based millimeter, MGS TES, and Viking atmospheric temperature measurements: Seasonal and interannual variability of temperatures and dust loading in the global Mars atmosphere. *J. Geophys. Res.*, 105, E4, doi: 10.1029/1999JE001089.

Clancy, R. T., M. J. Wolff, B. A. Whitney, B. A. Cantor, and M. D. Smith (2007), Mars equatorial mesospheric clouds: Global occurrence and physical properties from Mars Global Surveyor Thermal Emission Spectrometer and Mars Orbiter Camera limb observations, *J. Geophys. Res.*, 112, E04004, doi:10.1029/2006JE002805.

Colburn, D. S., J. B. Pollack, and R. M. Haberle (1989), Diurnal variations in optical depth at Mars, *Icarus*, **79**, 159-189, doi: 10.1016/0019-1035(89)90114-0.

Ellehoj, M. D., H.P. Gunnlaugsson, K.M. Bean*, B. A. Cantor, L. Drube, D. Fisher, B.T.Gheynani, A-M. Harri, C. Holstein-Rathlou, H. Kahanpää, M.T. Lemmon, M.B. Madsen, M. C. Malin, J. Polkko, P. Smith, L.K. Tamppari, P.A.Taylor, W. Weng and J. Whiteway (2010), Convective vortices and dust devils at the Phoenix Mars mission landing site, *J. Geophys. Res.*, 115, E00E16, doi:10.1029/2009JE003413.

Ferri, F., P. H. Smith, M. Lemmon, and N. O. Rennó (2003), Dust devils as observed by Mars Pathfinder, *J. Geophys. Res.*, 108(E12), 5133, doi:10.1029/2000JE001421.

Gierasch, Peter J., Richard M. Goody (1972). The effect of dust on the temperature of the Martian atmosphere. *J. Atmos. Sci.*, **29**, 400–402. doi: 10.1175/1520-0469(1972)029<0400:TEODOT>2.0.CO;2.

Greeley, R., D. A. Waller, N. A. Cabrol, G. A. Landis, M. T. Lemmon, L. D. V. Neakrase, M. Pendleton Hoffer, S. D. Thompson, and P. L. Whelley (2010), Gusev Crater, Mars:

Observations of three dust devil seasons, *J. Geophys. Res.*, 115, E00F02, doi:10.1029/2010JE003608.

Greeley, R., R.E. Arvidson, P.W. Barlett, D. Blaney, N.A. Cabrol, P.R. Christensen, R.L. Fergason, M.P. Golombek, G.A. Landis, M.T. Lemmon, S.M. McLennan, J.N. Maki, T. Michaels, J.E. Moersch, L.D.V. Neakrase, S.C.R. Rafkin, L. Richter, S.W. Squyres, P.A. de Souza, R.J. Sullivan, S.D. Thompson, and P.L. Whelley (2006a), Gusev crater: Wind-related features and processes observed by the Mars Exploration Rover Spirit, *J. Geophys. Res.*, 111, E02S09, doi:10.1029/2005JE002491.

Greeley, R., P.L. Whelley, R.E. Arvidson, N.A. Cabrol, D.J. Foley, B.J. Franklin, P.J. Geissler, M.P. Golumbeck, R.O. Kuzmin, G.A. Landis, M.T. Lemmon, L.D.V. Neakrase, S.W. Squyres, S. D. Thompson (2006b), Active dust devils in Gusev crater, Mars: Observations from the Mars Exploration Rover Spirit, *J. Geophys. Res.*, 111, E12S09, doi:10.1029/2006JE002743.Heavens, N. G., M. I. Richardson, A. Kleinböhl, D. M. Kass, D. J. McCleese, W. Abdou, J. L. Benson, J. T. Schofield, J. H. Shirley, and P. M. Wolkenberg (2011), The vertical distribution of dust in the Martian atmosphere during northern spring and summer: Observations by the Mars Climate Sounder and analysis of zonal average vertical dust profiles, *J. Geophys. Res.*, **116**, E04003, doi:10.1029/2010JE003691.

Hoekzema, N.M., M. Garcia-Comas, O.J. Stenzel, E.V. Petrova, N. Thomas, W.J. Markiewicz, K. Gwinner, H.U. Keller, and W.A. Delamere (2011). Retrieving optical depth from shadows in orbiter images of Mars. *Icarus* **214**, 447-461, doi: 10.1016/j.icarus.2011.06.009.

Hunt, G. E. (1979), On the opacity of Martian dust storms by Viking IRTM spectral measurements, *J. Geophys. Res.*, **84**, 8301–8310, doi: 10.1029/JB084iB14p08301.
41

Johnson, J. R., W. M. Grundy, M. T. Lemmon, J.F. Bell III, M.J. Johnson, R. Deen, R.E. Arvidson, W. Farrand, E. Guinness, K.E. Herkenhoff, F. Seelos IV, J. Soderblom, and S. Squyres (2006a), Spectrophotometric properties of materials observed by Pancam on the Mars Exploration Rovers: 1. Spirit, *J. Geophys. Res.*, **111**, E02S14, doi:10.1029/2005JE002494.

Johnson, J. R., W. M. Grundy, M. T. Lemmon, J.F. Bell III, M.J. Johnson, R. Deen, D. Alexander, R.E. Arvidson, W. Farrand, E. Guinness, A.G. Hayes, K.E. Herkenhoff, F. Seelos IV, J. Soderblom, and S. Squyres (2006b), Spectrophotometric properties of materials observed by Pancam on the Mars Exploration Rovers: 2. Opportunity, *J. Geophys. Res.*, **111**, E12S16, doi:10.1029/2006JE002762.

Johnson, J. R., W. M. Grundy, M. T. Lemmon (2003), Dust deposition at the Mars Pathfinder landing site: observations and modeling of visible/near-infrared spectra, *Icarus*, **163**, 330-346, doi: 10.1016/S0019-1035(03)00084-8.

Kinch, K. M., J. Sohl-Dickstein, J. F. Bell III, J. R. Johnson, W. Goetz, and G. A. Landis (2007), Dust deposition on the Mars Exploration Rover Panoramic Camera (Pancam) calibration targets, *J. Geophys. Res.*, 112, E06S03, doi:10.1029/2006JE002807.

Komguem, L., J.A. Whiteway, C. Dickinson, J.E. Moores, M. Day, J. Seabrook, and M.T. Lemmon (2013). Phoenix LIDAR Measurements of Mars Atmospheric Dust. *Icarus* **223**, 649-653. doi: 10.1016/j.icarus.2013.01.020.

Landis, G.A., and P.P. Jenkins (2000). Measurement of the settling rate of atmospheric dust on Mars by the MAE instrument on Mars Pathfinder. *J. Geophys. Res.*, **105**, 1855-1857, doi: 10.1029/1999JE001029.

Lemmon, M. T., M.J. Wolff, M.D. Smith, R.T. Clancy, D. Banfield, G.A. Landis, A. Ghosh, P.H. Smith, N. Spanovich, B. Whitney, P. Whelley, R. Greeley, S. Thompson, J.F. Bell

III, S.W. Squyres (2004), Atmospheric imaging results from the Mars Exploration Rovers: Spirit and Opportunity, *Science*, **306**, 1753–1756, doi: 10.1126/science.1104474..

Maki, J. N., J. F. Bell III, K. E. Herkenhoff, S. W. Squyres, A. Kiely, M. Klimesh, M. Schwochert, T. Litwin, R. Willson, A. Johnson, M. Maimone, E. Baumgartner, A. Collins, M. Wadsworth, S. T. Elliot, A. Dingizian, D. Brown, E. C. Hagerott, L. Scherr, R. Deen, D. Alexander, J. Lorre (2003). Mars Exploration Rover Engineering Cameras. *J. Geophys. Res.*, **108**, E12, 8071, doi:10.1029/2003JE002077.

Malin, M.C., and K.E. Edgett (2001). Mars Global Surveyor Mars Orbiter Camera: Interplanetary cruise through primary mission. *J. Geophys. Res.*, **106**, E10, 23429-23570, doi: 10.1029/2000JE001455.

Malin, M.C., K.S. Edgett, B.A. Cantor, M.A. Caplinger, G.E. Danielson, E.H. Jensen, M.A. Ravine, J. L. Sandoval, and K.D. Supulver (2010). An overview of the 1985-2006 Mars Orbiter Camera science investigation. *Mars* 5, 1-60, 2010, doi:10.1555/mars.2010.0001.

Martin, T. Z. (1986), Thermal infrared opacity of the Mars atmosphere, *Icarus*, **66**, 2–21

Mateshvili, N., D. Fussen, F. Vanhellemont, C. Bingen, J. Dodion, F. Montmessin, S. Perrier, J.L. Bertaux (2007), Detection of Martian dust clouds by SPICAM UV nadir measurements during the October 2005 regional dust storm, *Adv. Space Res.*, **40**, Issue 6, 869-880, doi: 10.1016/j.asr.2007.06.028.

Metzger, S. M., J. R. Carr, J. R. Johnson, T. J. Parker, and M. T. Lemmon (1999), Dust devil vortices seen by the Mars Pathfinder Camera, *Geophys. Res. Lett.*, **26**(18), 2781–2784, doi:10.1029/1999GL008341.

Montabone, L., F. Forget, E. Millour, R. J. Wilson, S. R. Lewis, D. Kass, A. Kleinboehl, M. T. Lemmon, M. D. Smith, and M. J. Wolff. Eight-year Climatology of Dust on Mars. *Icarus*, this issue.

Newman, C. E., S. R. Lewis, P. L. Read, and F. Forget (2002), Modeling the Martian dust cycle 2. Multiannual radiatively active dust transport simulations, *J. Geophys. Res.*, **107**(E12), 5124, doi:10.1029/2002JE001920.

Petrova, E.V., N.M. Hoekzema, W.J. Markiewicz, N. Thomas, and O.J. Stenzel (2012). Optical depth of the Martian atmosphere and surface albedo from high-resolution orbiter images. *Planet. Space Sci*. **60**, 287-296, doi: 10.1016/j.pss.2011.09.008.

Pollack, J. B., D. S. Colburn, F. M. Flasar, R. Kahn, C. E. Carlston, and D. G. Pidek (1979), Properties and effects of dust particles suspended in the Martian atmosphere, *J. Geophys. Res.*, **84**, 2929–2945. doi: 10.1029/JB084iB06p02929.

Pollack, J. B., M.E. Ockert-Bell, M.K. Shepard (1995). Viking Lander image analysis of Martian atmospheric dust. *J. Geophys. Res.*, **100**(E3), 5235-5250, doi:10.1029/94JE02640.

Rafkin, S. C. R. (2009), A positive radiative-dynamic feedback mechanism for the maintenance and growth of Martian dust storms, *J. Geophys. Res.*, 114, E01009, doi:10.1029/2008JE003217.

Rice, M. S., J. F. Bell III, E. A. Cloutis, J. J. Wray, K. E. Herkenhoff, R. Sullivan, J. R. Johnson, and R. B. Anderson (2011), Temporal observations of bright soil exposures at Gusev crater, Mars, *J. Geophys. Res.*, 116, E00F14, doi:10.1029/2010JE003683.

Schuerger, A. C., J.T. Richards, D.A. Newcombe, K. Venkateswaran (2006), Rapid inactivation of seven Bacillus spp. under simulated Mars UV irradiation, *Icarus*, **181**, 52-62, doi: 10.1016/j.icarus.2005.10.008.

Schuerger, A. C., P. Fajardo-Cavazos, C.A. Clausen, J.E. Moores, P. H. Smith, W.L. Nicholson (2008), Slow degradation of ATP in simulated martian environments suggests long residence times for the biosignature molecule on spacecraft surfaces on Mars, *Icarus*, **194**, 86-100, doi: 10.1016/j.icarus.2007.10.010.

Shaw, A., M.J. Wolff, F.P. Seelos, S.M. Wiseman, and S. Cull (2013). Surface scattering properties at the Opportunity Mars rover's traverse region measured by CRISM. *J. Geophys. Res.*, 118, 1699-1717, doi: 10.1002/jgre.20119.

Sirianni, M., M. J. Jee, N. Benítez, J. P. Blakeslee, A. R. Martel, G. Meurer, M. Clampin, G. De Marchi, H. C. Ford, R. Gilliland, G. F. Hartig, G. D. Illingworth, J. Mack, and W. J. McCann. The Photometric Performance and Calibration of the Hubble Space Telescope Advanced Camera for Surveys. *Publications of the Astronomical Society of the Pacific*, **117**, No. 836 (October 2005), 1049-1112, DOI: 10.1086/444553.

Smith, M. D. (2004). Interannual variability in TES atmospheric observations of Mars during 1999-2003. *Icarus,* **167**, 148-165, doi: 10.1016/j.icarus.2003.09.010.

Smith, M. D. (2008). Spacecraft observations of the Martian atmosphere. *Ann. Rev. Earth Planet. Sci.*, **36**, 191-219, doi: 10.1146/annurev.earth.36.031207.124334.

Smith, M. D. (2009). THEMIS observations of Mars aerosol optical depth from 2002-2008. *Icarus,* **202**, 444-452, doi: 10.1016/j.icarus.2009.03.027.

Smith, M. D., M. J. Wolff, N. Spanovich, A. Ghosh, D. Banfield, P. R. Christensen, G. A. Landis, and S. W. Squyres (2006), One Martian year of atmospheric observations using MER Mini-TES. *J. Geophys. Res.,* **111**, E12S13, doi:10.1029/2006JE002770.

Smith, M. D., M.J. Wolff, M.T. Lemmon, N. Spanovich, D. Banfield, C.J. Budney, R.T. Clancy, A. Ghosh, G.A. Landis, P. Smith, B. Whitney, P.R. Christensen, and S.W. Squyres, 2004. First atmospheric science results from the Mars Exploration Rovers Mini-TES. *Science* **306**, 1750-1753, doi: 10.1126/science.1104257.

Smith, P. H., and M. Lemmon (1999), Opacity of the Martian atmosphere measured by the Imager for Mars Pathfinder, *J. Geophys. Res.*, 104(E4), 8975–8985, doi:10.1029/1998JE900017.

Squyres, S. W., R.E. Arvidson, D. Bollen, J.F. Bell, J. Brückner, N.A. Cabrol, W.M. Calvin, M.H. Carr, P.R. Christensen, B.C. Clark, L. Crumpler, D.J. Des Marais, C. d'Uston, T. Economou, J. Farmer, W.H. Farrand, W. Folkner, R. Gellert, T.D. Glotch, M. Golombek, S. Gorevan, J.A. Grant, R. Greeley, J. Grotzinger, K.E. Herkenhoff, S. Hviid, J.R. Johnson, G. Klingelhöfer, A.H. Knoll, G. Landis, M. Lemmon, R. Li, M.B. Madsen, M.C. Malin, S.M. McLennan, H.Y. McSween, D.W. Ming, J. Moersch, R.V. Morris, T. Parker, J.W. Rice, L. Richter, R. Rieder, C. Schröder, M. Sims, M. Smith, P. Smith, L.A. Soderblom, R. Sullivan, N.J. Tosca, H. Wänke, T. Wdowiak, M. Wolff, and A. Yen (2006), Overview of the Opportunity Mars Exploration Rover Mission to Meridiani Planum: Eagle Crater to Purgatory Ripple, *J. Geophys. Res.*, 111, E12S12, doi:10.1029/2006JE002771.

Tomasko, M. G., L. R. Doose, M. Lemmon, P. H. Smith, and E. Wegryn (1999), Properties of dust in the Martian atmosphere from the Imager on Mars Pathfinder, *J. Geophys. Res.*, **104**, 8987–9008, doi:10.1029/1998JE900016.

Vincendon, M., F. Forget, and J. Mustard (2010), Water ice at low to midlatitudes on Mars, *J. Geophys. Res.*, 115, E10001, doi:10.1029/2010JE003584.

Vincendon, M., Y. Langevin, F. Poulet, A. Pommerol, M. Wolff, J.-P. Bibring, B. Gondet, D. Jouglet (2009), Yearly and seasonal variations of low albedo surfaces on Mars in the OMEGA/MEx dataset: Constraints on aerosols properties and dust deposits, *Icarus*, **200**, 395-405, doi: 10.1016/j.icarus.2008.12.012.

Vincendon, M., Y. Langevin, F. Poulet, J.-P. Bibring, and B. Gondet (2007), Recovery of surface reflectance spectra and evaluation of the optical depth of aerosols in the near-IR using a Monte Carlo approach: Application to the OMEGA observations of high-latitude regions of Mars, *J. Geophys. Res.*, **112**, E08S13, doi:10.1029/2006JE002845.

Wang, H., M. I. Richardson, R. J. Wilson, A. P. Ingersoll, A. D. Toigo (2003), Cyclones, tides, and the origin of a cross-equatorial dust storm on Mars, *Geophys. Res. Lett.*, **30**(9), 1488, doi:10.129/2002GL016828.

Wang, H., (2007). Dust storms originating in the northern hemisphere during the third mapping year of Mars Global Surveyor. *Icarus* **189**, 325-343. doi: 10.1016/j.icarus.2007.01.014.

Whiteway, J. A., L. Komguem, C. Dickinson, C. Cook, M. Illnicki, J. Seabrook, V. Popovici, T. J. Duck, R. Davy, P. A. Taylor, J. Pathak, D. Fisher, A. I. Carswell, M. Daly, V. Hipkin, A.P. Zent, M.H. Hecht, S.E. Wood, L. Tamppari, N. Renno, J. Moores, M.T. Lemmon, F. Daerden, P.H. Smith, (2009). Mars Water Ice Clouds and Precipitation. *Science* **325**, 68-70. doi: 10.1126/science.1172344.

Wolff, M. J., M.D. Smith, R.T. Clancy, N. Spanovich, B.A. Whitney, M.T. Lemmon, J. Bandfield, D. Banfield, A. Ghosh, G. Landis, P. Christensen, J.F. Bell III, and S. Squyres (2006), Constraints on dust aerosols from the Mars Exploration Rovers using MGS overflights and Mini-TES, *J. Geophys. Res.*, **111**, E12S17, doi:10.1029/2006JE002786.

Wolff, M. J., M. D. Smith, R. T. Clancy, R. Arvidson, M. Kahre, F. Seelos IV, S. Murchie, and H. Savijärvi (2009), Wavelength dependence of dust aerosol single scattering albedo as observed by the Compact Reconnaissance Imaging Spectrometer, *J. Geophys. Res.*, **114**, E00D04, doi:10.1029/2009JE003350,.

Wolff,, M. J., R. T. Clancy, J.D. Goguen, M.C. Malin, B.A. Cantor (2010), Ultraviolet dust aerosol properties as observed by MARCI, *Icarus*, **208**, 143-155, doi: 10.1016/j.icarus.2010.01.010.

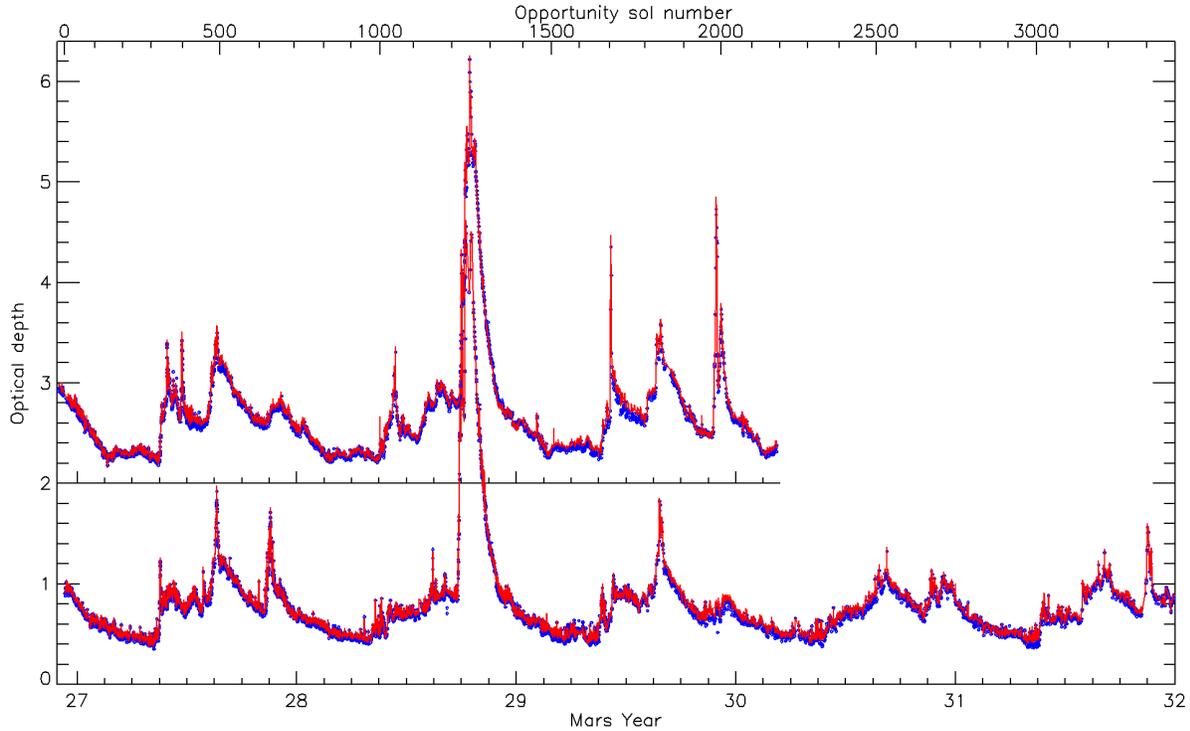

Figure 1. The MER optical depth record for Opportunity and Spirit (offset by 2). Opacity observed in L8 (blue symbols) and R8 (red line) is shown. Mars year is 27.0 at northern spring equinox ($L_S=0°$, Opportunity sol 41) following landing and increases proportionally to $L_S$ between successive northern spring equinoxes; minor ticks indicate 45° of $L_S$. A sol scale along the top axis indicates Opportunity sol number (20.5 less than Spirit sol number), which is non-linear in $L_S$. All valid measurements are presented.

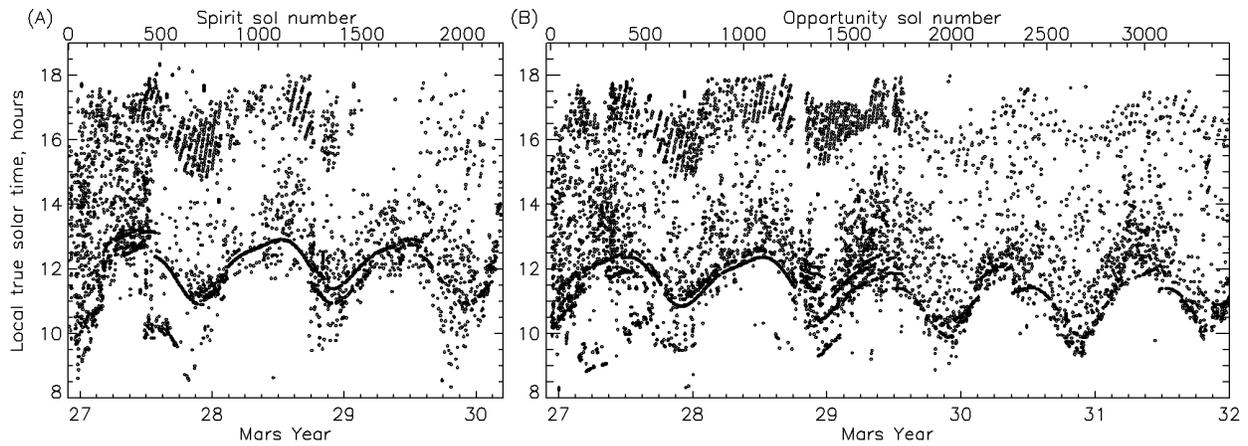

Figure 2. Local true solar time (LTST) for each optical depth measurement for (A) Spirit and (B) Opportunity. LTST midnight is defined as 0h and 24h when the Sun is at its lowest ephemeris position and increases linearly with time through a sol. The pattern near 12h is associated with the direct to/from Earth communication; patterns near 16h are associated with Odyssey relay communications.

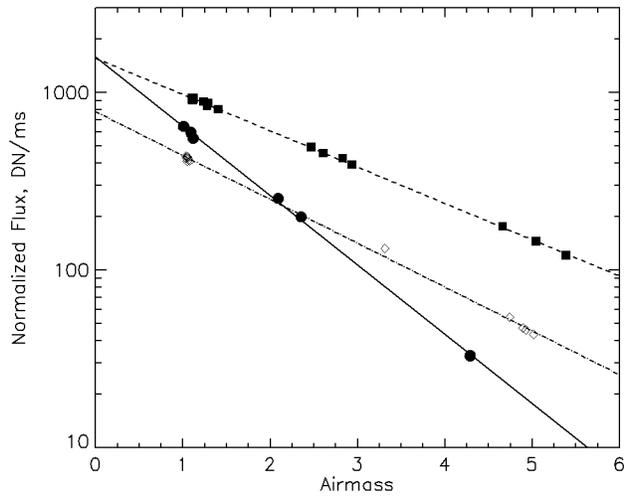

Figure 3. Beer-Lambert Law fit. Opportunity L8 fluxes, normalized to a Sun-Mars distance of 1.50 astronomical units, from sols 23 (filled circles), 179-184 (filled squares), and 1691-1694 (open diamonds) are shown as a function of airmass along with exponential fits for each period. The early data has nearly the same 0-airmass intercept, but different opacity (0.90, then 0.47). The post-dusting data have a moderate atmospheric opacity (0.57) plus an additional 50% extinction from window-dust. Uncertainties are smaller than the symbols.

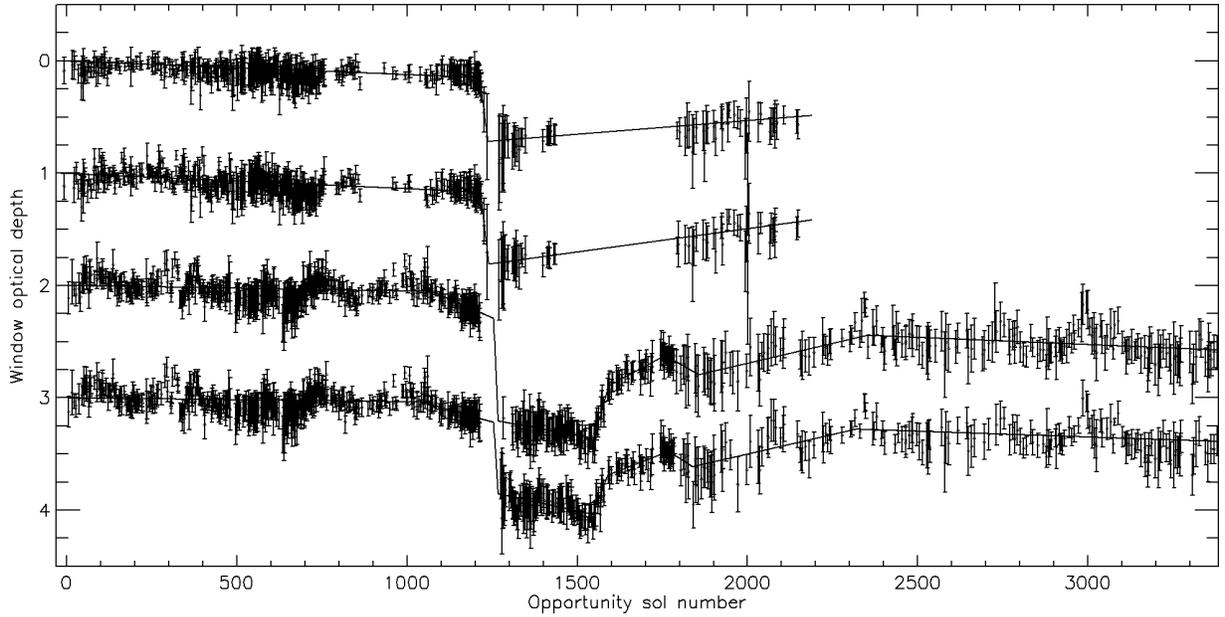

Figure 4. Accumulation of window dust. Beer-Lambert law intercepts for individual sols are shown for each rover and filter, with all changes from sol 1 interpreted as accumulated opacity on the Pancam windows, which are physically separated for left and right eyes. Spirit L8 fits are at the top; Spirit R8, Opportunity L8 and R8 are successively offset by 1 unit optical depth. The abscissa is Opportunity sol number (Spirit sols are offset from Opportunity by -20.5). The ordinate is reversed to show accumulated optical depth as an absorption such that the Beer-Lambert intercept for a given sol is depressed: $\ln(F_{sol}/F_{TOA}) = \tau$. The window dust model is a 3-part line for Spirit (accumulation, rapid accumulation, removal) and an 8 part line for Opportunity.

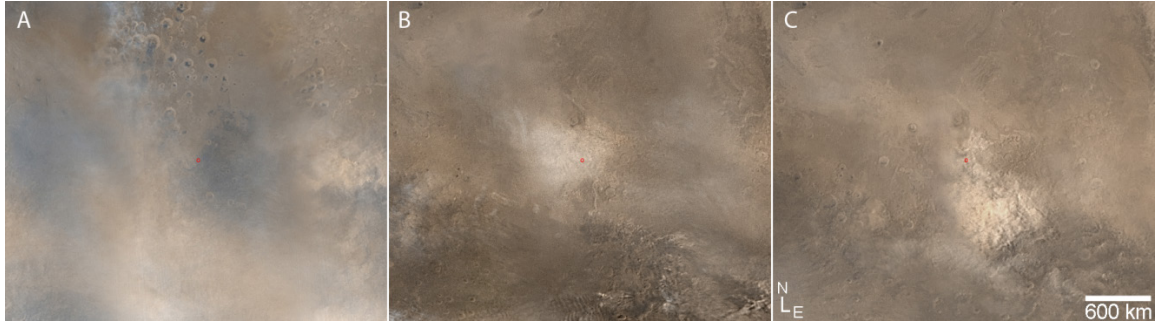

Figure 5. Local dust storm activity at the two MER sites. (A) In Meridiani Planum on sol B/330 (28 December 2004, $L_s$=135°) the optical depth was near its peak of 1.2. (B) In Gusev Crater on sol A/372 (19 January 2005, $L_s$=147°) the optical depth was 1.1, and peaked at 1.4 the next sol, with slightly elevated optical depths after the 2-sol event. (C) In Gusev crater on sol A/420 (9 March 2005, $L_s$=173°) the optical depth peaked at 1.4 in the afternoon, and varied from 0.7-1.5 within a 10-sol interval. Maps are composed of MOC daily global mapping images that have been cylindrically map projected at 6 km pixel$^{-1}$ and mosaicked together. The color has been stretched to enhance the contrast between the storms and the Martian surface. The Meridiani frame spans 27.2°S-22.8°N, 154.0°-214.0°W and the two Gusev frames span 40.0°S-10.0°N, 155.0°-215.0°W. The small circle near the center of each image frame corresponds to the location of the rovers.

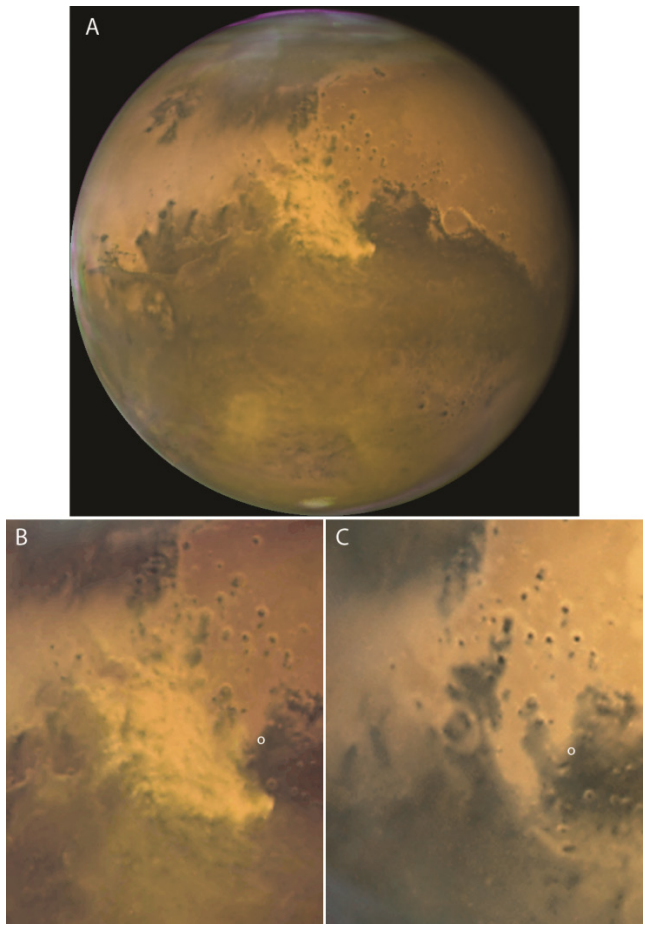

Figure 6. Regional dust storm activity entering the Meridiani region. (A) A global view was acquired by the Hubble Space Telescope's Advanced Camera for Surveys, using 250-, 502-, and 658-nm filters, on 28 October 2005 ($L_S=314°$, sol B/626). The area of the 1500-km long storm is shown (B) in a 2500-km wide excerpt of the global view and (C) in a comparable Wide Field Planetary Camera 2 image acquired 26 June 2001 with storm-free conditions. The circle shows the Opportunity site, where the optical depth was 1.4 and peaked at 1.8, 4 sols later. [Image credit: NASA, ESA, The Hubble Heritage Team (STScI/AURA), J. Bell (Cornell University) and M. Wolff (Space Science Institute).]

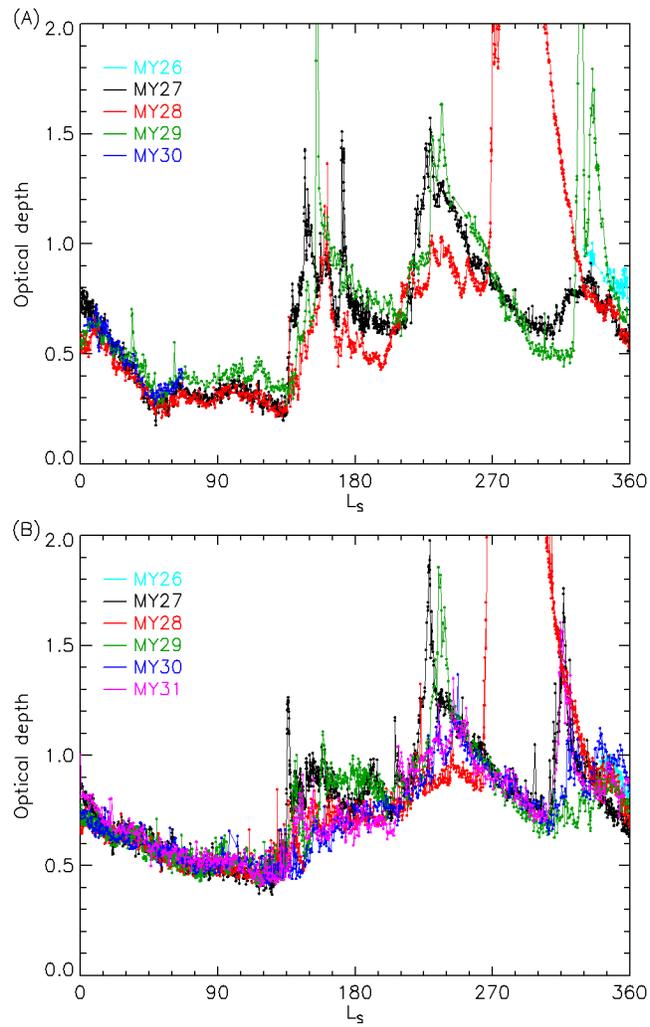

Figure 7. 880-nm optical depths as a function of season. Shown are results from (A) Spirit and (B) Opportunity for Mars years 26 (cyan, $L_S>330$ only), 27 (black), 28 (red), 29 (green), 30 (blue), and 31 (magenta). The 2007 dust storm is cut off to show detail at other times.

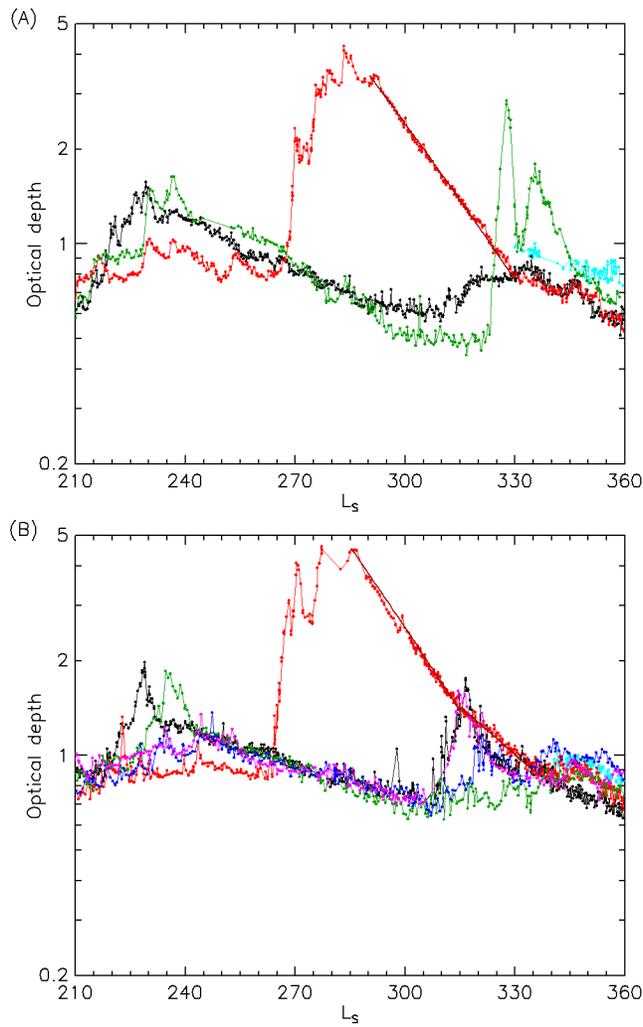

Figure 8. 880-nm optical depths for southern summer and autumn. Conventions are the same as Fig. 7, except the logarithmic axis shows the full range of optical depths. Solid maroon lines illustrate (in A) a 2.3%/sol decay over $L_S=290°$-$330°$, and (in B) a 2.4%/sol decay over $L_S=285°$-$315°$ and a 1.3%/sol decay over $L_S=315°$-$335°$.

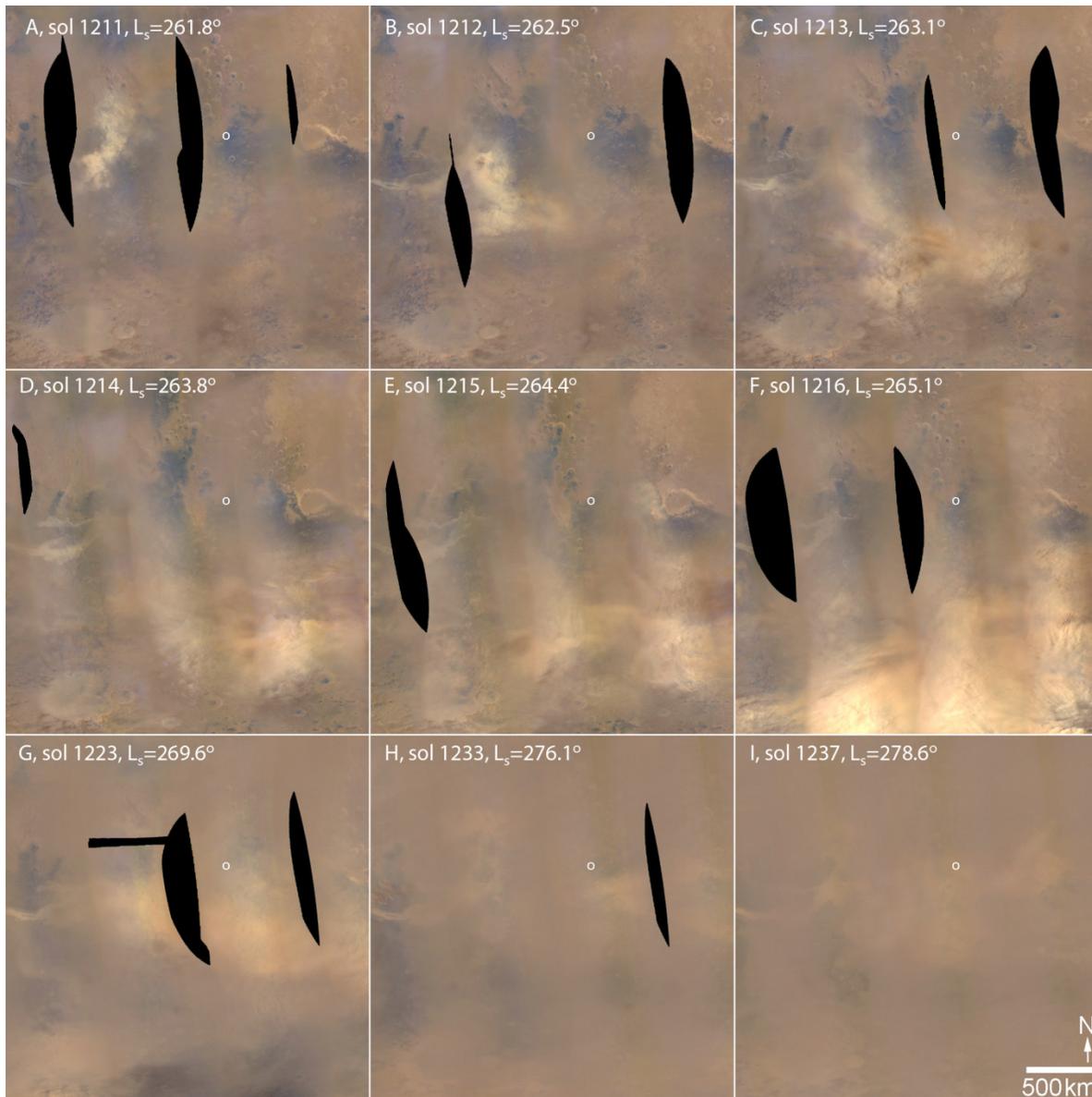

Figure 9. The 2007 global dust event at the Opportunity site. Shown are excerpts, spanning from 60°S-30°N, 330°-60°W, from Mars Reconnaissance Orbiter Mars Color Imager daily mosaics over 21 June to 18 July 2007 (sols B/1211-1237, L$_S$=262°-279°), at a (summed-8) resolution of 7.5 km/pixel. The white circle in each image shows the Opportunity site; black areas are gaps in data. During 21-26 June (A-F), atmospheric optical depth increased from 0.9 to 1.4 and became variable on time scales <1 sol. (G) The atmospheric optical depth was 2.8 on sol 1223, 3 July; (H) 3.5-3.9 (variable) on sol 1233, 14 July; and (I) unmeasured on sol 1237, 18 July. During sols 1236-1241, the rover was shut down, but solar panel output suggests a peak optical depth near 5 on sol 1237.

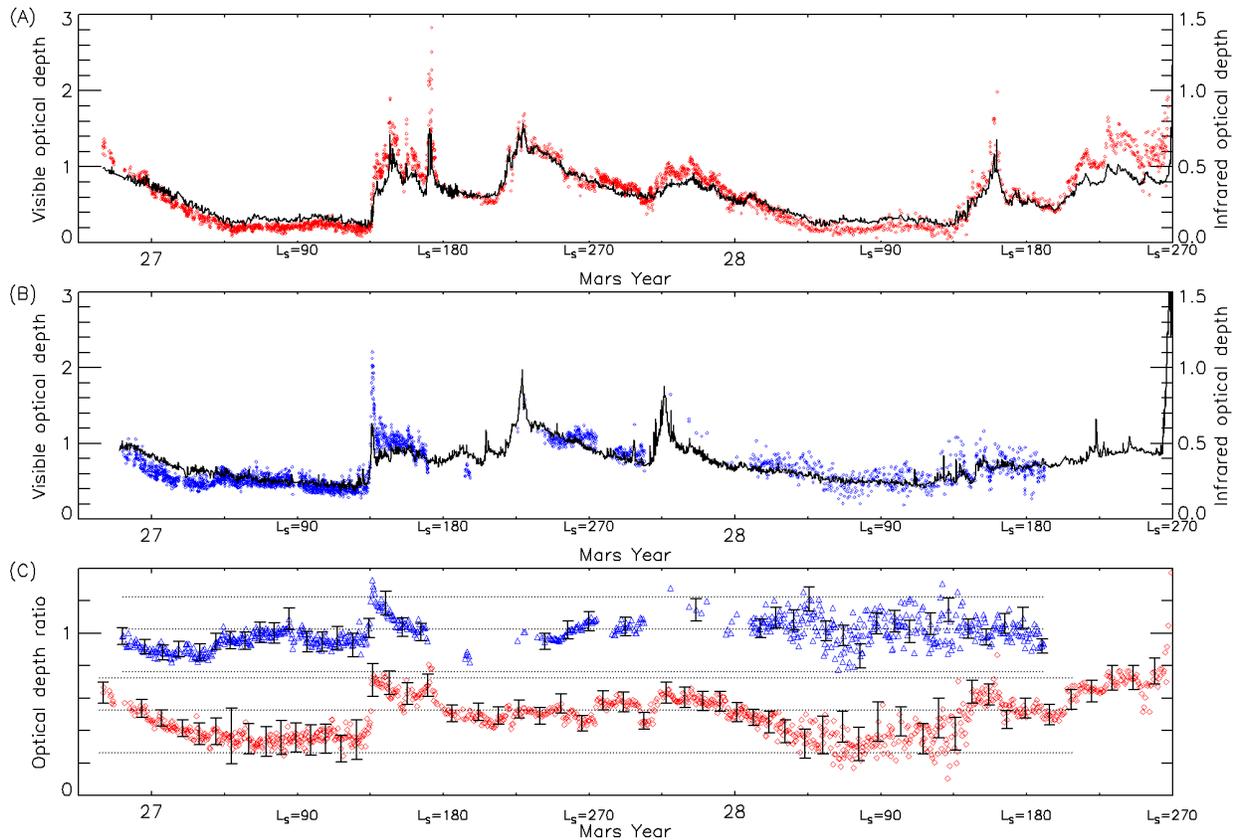

Figure 10. Visible to infrared opacity comparison. (A) The 0.88-μm (R8) opacity (black line) is compared with dust opacity observed by MiniTES at 9-μm (red diamonds, see Smith *et al.* [2006]). Uncertainties, not shown, are typically small compared to the symbols. (B) The same is shown for Opportunity, in blue. (C) The ratio of MiniTES to R8 opacity is shown; every 20$^{th}$ error bar is plotted. Opportunity values are offset by 0.5. The dotted lines correspond to aerosol size distributions with variance 0.5 and mean radius of 0.7, 1.4, and 2.1 μm, from bottom (see text).

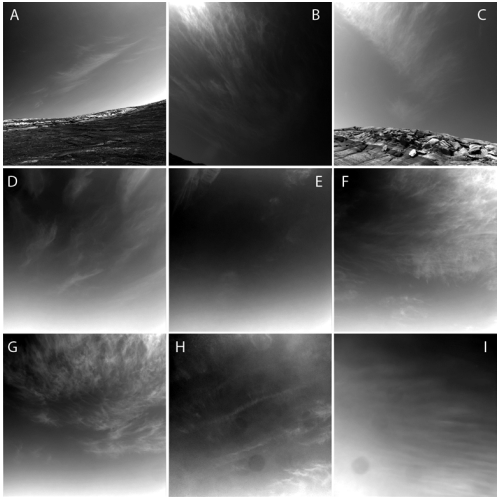

Figure 11. Opportunity Navcam images of clouds acquired on sol (A) 269, $L_S$=106°; (B) 290, 116°; (C) 291, 117°; (D) 758, 24°; (E) 763, 27°; (F) 949, 112°; (G) 950, 112°; (H) 1647, 126°; and (I) 2847, 63°. All images have a linear contrast-stretch; clouds are typically near 15% contrast at their peak, except for (B) 30% and (I) 5%.

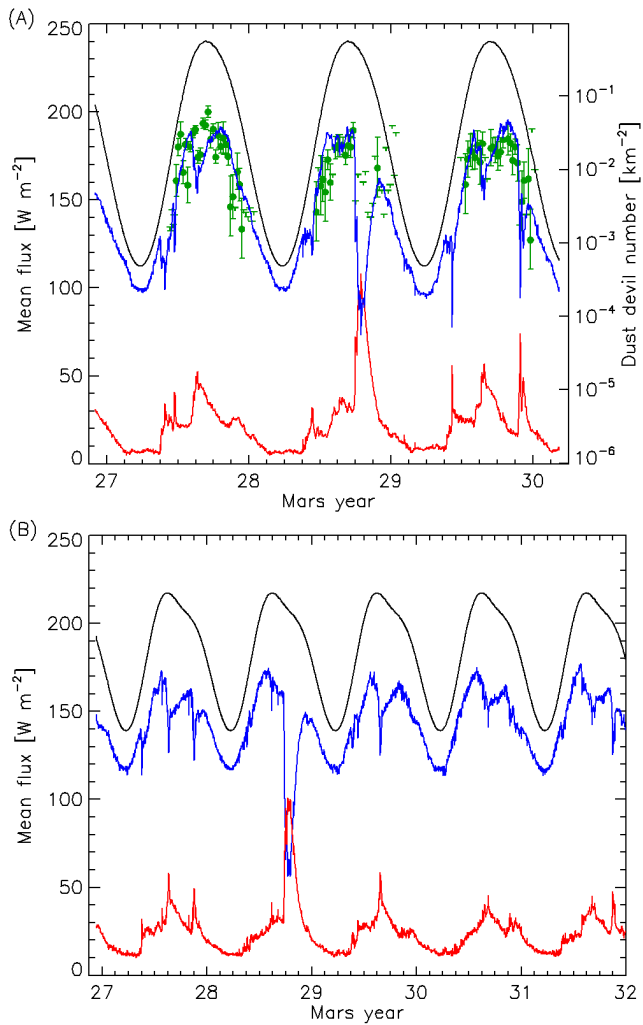

Figure 12. Insolation at the rover sites for (A) Spirit and (B) Opportunity. The continuous curves show modeled top-of-atmosphere, sol-average insolation (upper, black), direct plus diffuse surface insolation (middle, blue) and atmospheric absorption of sunlight (lower, red). For Spirit, symbols (green) show dust devil number density (right axis) reported by Greeley *et al.* [2010] with ┬ symbols indicating upper limits.